\documentclass[12pt]{article}
\usepackage{graphics, color}
\usepackage{amsthm}
\usepackage{amsmath}
\usepackage{amsfonts}
\usepackage[dvips]{graphicx}
\usepackage{amssymb}
\def\pplogo{\vbox{\kern-\headheight\kern -29pt
\halign{##&##\hfil\cr&{\ppnumber}\cr\rule{0pt}{2.5ex}&\ppdate\cr}}}
\makeatletter
\def\ps@firstpage{\ps@empty \def\@oddhead{\hss\pplogo}%
  \let\@evenhead\@oddhead 
}
\def\maketitle{\par
 \begingroup
 \def\thefootnote{\fnsymbol{footnote}}
 \def\@makefnmark{\hbox{$^{\@thefnmark}$\hss}}
 \if@twocolumn
 \twocolumn[\@maketitle]
 \else \newpage
 \global\@topnum\z@ \@maketitle \fi\thispagestyle{firstpage}\@thanks
 \endgroup
 \setcounter{footnote}{0}
 \let\maketitle\relax
 \let\@maketitle\relax
 \gdef\@thanks{}\gdef\@author{}\gdef\@title{}\let\thanks\relax}
\makeatother

\numberwithin{equation}{section}

\renewcommand{\dag}{\dagger}

\newcommand{\OO}{\mathcal{O}}

\newcommand{\be}{\begin{eqnarray}}
\newcommand{\bea}{\begin{eqnarray}}
\newcommand{\ee}{\end{eqnarray}}
\newcommand{\eea}{\end{eqnarray}}

\newcommand{\f}{\frac}

\newcommand{\Tr}{{\rm Tr}}
\newcommand{\tr}{{\rm tr}}
\renewcommand{\t}{\tilde}

\newcommand{\Qt}{\tilde{Q}}
\newcommand{\muphi}{\mu_\phi}
\newcommand{\Ntilde}{\tilde{N}}

\newcommand{\Zt}{\tilde{Z}}
\newcommand{\chit}{\tilde{\chi}}
\newcommand{\rhot}{\tilde{\rho}}

\newcommand{\Nct}{\tilde{N}_c}

\newcommand{\ZUt}{\tilde{Z}_U}
\newcommand{\ZU}{Z_U}
\begin{document}

\setcounter{page}0
\def\ppnumber{\vbox{\baselineskip14pt
}}
\def\ppdate{\footnotesize{SLAC-PUB-13838, SU-ITP-09-50, NSF-KITP-09-199}} \date{}

\author{Nathaniel Craig$^{1}$, Rouven Essig$^{1}$, Sebasti\'an Franco$^2$,\\ Shamit Kachru$^{2}$\footnote{On leave from Department of Physics and SLAC, Stanford University.}, Gonzalo Torroba$^{1,\,2}$\\
[7mm]
{\normalsize $^1$ Physics Department and SLAC National Accelerator Laboratory, }\\
{\normalsize Stanford University, Stanford, CA 94309, USA}\\
{\normalsize $^2$ Kavli Institute for Theoretical Physics and Physics Department, }\\
{\normalsize University of California, Santa Barbara, CA 93106--4030, USA}\\
}

\title{\bf Dynamical Supersymmetry Breaking,\\
[2mm]
with Flavor
\vskip 0.4cm}
\maketitle

\begin{abstract} \normalsize
\noindent

We explore calculable models with low-energy supersymmetry where the flavor 
hierarchy is generated by quark and lepton compositeness, and where
the composites emerge from the same sector that dynamically breaks supersymmetry. 
The observed pattern of Standard Model fermion masses and mixings is obtained by 
identifying the various generations with composites of different dimension in the ultraviolet. 
These ``single-sector'' supersymmetry breaking models give rise to various spectra of soft 
masses which are, in many cases, quite distinct from what is commonly found in models of 
gauge or gravity mediation. In typical models which satisfy all flavor-changing neutral current 
constraints, both the first and second generation sparticles have masses of order 20 TeV, while
the stop mass is a few TeV. In other cases, all sparticles obtain masses of order a few TeV 
predominantly from gauge mediation, even though the first two generations are composite.

\end{abstract}
\bigskip
\newpage

\tableofcontents

\vskip 1cm

\section{Introduction}\label{sec:intro}

Two central mysteries in fundamental physics involve the discrepancy between $G_{\rm Fermi}$
 and $G_{\rm Newton}$, and the origin of the patterns in the quark and lepton Yukawa couplings.  Supersymmetry is
a well motivated candidate which addresses the first question.  It is then natural to ask, can we find
supersymmetric models of weak scale physics where both questions are answered simultaneously, and
the dynamics that explains the weak scale also explains the texture of the fermion mass matrix?

One promising idea which could explain the structure of the Yukawa couplings is compositeness.  If the first two generations
of quarks and leptons are composites at some intermediate scale $\Lambda$, while flavor physics is 
generated at $M_{\rm flavor} \gg \Lambda$, then the masses and mixings of the first two generations will
be suppressed by the small parameter $\epsilon \equiv \Lambda / M_{\rm flavor}$.  
The third generation should be elementary (external to the strong dynamics), because the top quark Yukawa 
coupling is $\OO(1)$ and thus not suppressed. 
It was proposed in
\cite{ArkaniHamed:1997fq, Luty:1998vr} that perhaps the strongly-coupled sector that is
responsible for dynamical supersymmetry breaking could also generate the first two generations of
quarks and leptons as composites of the same strong dynamics.  Such ``single-sector"
models could give a simultaneous
explanation of the Planck/weak hierarchy and the masses and mixings of Standard Model particles.

While this is an attractive idea, there were no calculable examples.  Recently, using the fact that 
supersymmetric QCD (SQCD) has simple metastable vacua that exhibit dynamical supersymmetry breaking
\cite{ISS}, calculable examples of such single-sector models were developed \cite{Franco:2009wf}.  The simplest examples
give rise to two composite generations, both arising from dimension two operators in the high energy
theory.  The natural texture of the matrix of masses and mixings is then of the form
\be
\label{dimtwo}
\left( \begin{array}{ccc}
\epsilon^2 & \epsilon^2 & \epsilon \\
\epsilon^2 &\epsilon^2 & \epsilon      \\
\epsilon & \epsilon    & 1 
\end{array}
\right)~.
\ee
In the models of \cite{Franco:2009wf}, the first two generations of sparticles are parametrically heavier than the third generation sparticles.

It is desirable, however, to find other classes of calculable single-sector models where the
mass matrix can take a more general form.  For instance, if one of the generations arises from a
dimension three operator in the high energy theory, while the other arises from a dimension two
operator, one would expect a mass matrix of the slightly more appealing form
\be
\label{simple}
\left( \begin{array}{ccc}
\epsilon^4 & \epsilon^3 & \epsilon^2 \\
\epsilon^3 &\epsilon^2 & \epsilon      \\
\epsilon^2 & \epsilon    & 1 
\end{array}
\right)~.
\ee
With additional $\OO(1)$ coefficients and $\epsilon \sim 0.1$, this Yukawa matrix reproduces the observed 
flavor hierarchy. 

Our goal in this paper is to explore the class of calculable single-sector models that can be
constructed given the current state-of-the-art in models of dynamical supersymmetry breaking.
We will find that models with this flavor structure --- as well as models with additional parameters that
give more general classes of mass matrices --- can easily be constructed.

In the models of \cite{ArkaniHamed:1997fq, Luty:1998vr}, as well as the newer calculable models in
\cite{Franco:2009wf}, the composite generations not surprisingly couple more strongly to the
supersymmetry-breaking order parameter than the elementary third generation (whose leading sfermion mass arises
from gauge mediation, after weakly gauging the Standard Model subgroup of the global symmetry
group of the supersymmetry-breaking theory).  Therefore, one is led to phenomenology very
reminiscent of the scenario advocated in \cite{Dimopoulos:1995mi,Cohen:1996vb}, where the 
first and second generation sfermion masses are larger than 
those of the third generation.  One of the surprises we shall find here is that in some of our models even
some of the composites can have leading masses arising from gauge mediation and comparable
to the third generation masses.

\subsection{General strategy}\label{subsec:strategy}

Before we proceed to a detailed analysis, it is worth explaining the general strategy.  One of the most elegant ideas for explaining the texture of Yukawas given by Eq.~(\ref{simple}), which matches observation reasonably well, is to postulate that the first and second generations are
secretly composite above some scale $\Lambda$, and in the high-energy theory their Yukawa couplings
are then irrelevant operators.  With a first and second generation emerging from operators whose
dimensions in the ultraviolet (UV) 
are 3 and 2 (and an elementary third generation), one naturally gets  the structure above, with the small parameter
\be
\epsilon = \Lambda / M_{\rm flavor}
\ee
emerging from the suppression of irrelevant operators in the high-energy theory.  For $\epsilon \sim 10^{-1}$,
this is an excellent starting point for matching observations.

More concretely, consider an asymptotically free SQCD theory with gauge 
group $G$, fundamental quarks $(Q, \t Q)$ and a field $U$ in a 2-index tensor 
representation of the gauge group. We will call this the ``electric theory'', and its dynamical scale, below which it becomes 
strongly coupled, will be denoted by $\Lambda$. 

A promising approach to constructing calculable models arises when the theory 
has an infrared dual description (the ``magnetic theory'') where the mesons 
$(Q U \t Q)$ and $(Q \t Q)$ are weakly coupled. 
These are the fields that will produce the first and second generations. Generically, the IR theory 
also contains magnetic quarks $(q, \t q)$, and a field $\t U$ in a rank 2 tensor representation of the magnetic gauge group.

Furthermore, we imagine that there is some additional UV physics at a scale $M_{\rm flavor} > \Lambda$, responsible 
for generating the Yukawa couplings\footnote{The MSSM contains separate $H_u$ and $H_d$ fields, but we will 
simplify schematic equations of this sort by just denoting both Higgs fields by $H$ throughout the paper.}
\be
W_{Yuk} & \supset & \frac{1}{M_{\rm flavor}^4} (QU\t Q) H (QU\t Q) + \frac{1}{M_{\rm flavor}^3} (Q\t Q) H (QU \t Q) + \nonumber \\
& & \f{1}{M_{\rm flavor}^2} (Q\t Q) H (Q\t Q) + \f{1}{M_{\rm flavor}} (Q\t Q) H \Psi_3 + \Psi_3 H \Psi_3~.
\ee
Here $\Psi_3$ denotes the elementary third generation. 
Rescaling the fields by appropriate powers of $\Lambda$ so that they are canonically normalized gives a Yukawa matrix of the form (\ref{simple}). 

In general, the mesons $(Q \t Q)$ and $(Q U \t Q)$ contain more matter than just the first two Standard Model generations. It will be shown 
that some of the extra components of these fields together with the magnetic quarks yield a weakly coupled supersymmetry breaking model 
(as in~\cite{ISS}). In this effective description, supersymmetry breaking occurs through tree-level and one-loop interactions, while 
the supersymmetry breaking scale is generically an inverse loop factor above the electroweak scale.  

\vskip 4mm

The organization of the paper is as follows.  In \S \ref{sec:simple}, we present the simplest model which naturally
gives rise to two composite generations with a Yukawa matrix of more general type than (\ref{dimtwo}).  This model has two parameters in the flavor sector instead of one, and so while it can model observations quite well, it is perhaps less elegant than the more predictive structure in 
(\ref{simple}).  Therefore, in \S \ref{sec:adjoint}, we move on to a class of models which give rise to the structure (\ref{simple}).  A starring role is played by the metastable supersymmetry-breaking vacua of 
SQCD with fundamental flavors and an additional adjoint chiral superfield. After discussing the asymptotically free electric theory and its infrared free magnetic dual, we find new metastable supersymmetry breaking vacua.

In \S \ref{sec:adj-ssector}, we show how this simple model in \S\ref{sec:adjoint} naturally explains 
the flavor hierarchy and present the fermion and 
sparticle spectrum.  We also discuss constraints on the sparticle spectra from flavor changing neutral currents (FCNCs). The simplest model is consistent with the constraints from FCNCs only 
in a small region of parameter space, and in \S\ref{sec:sols}, we present more general models that accommodate current bounds.

We present our conclusions in \S \ref{sec:conclusions},  
where we also briefly compare this method of explaining the Yukawa flavor pattern to other
common explanations in the literature.  Two appendices are devoted to a more careful discussion of FCNCs (Appendix \ref{sec:FCNC}) and a discussion of gauge coupling unification and the existence of Landau poles (Appendix \ref{sec:unification}). 
Since all of the models we study will typically have a lot of extra massive matter at very high scales, gauge coupling 
unification can be challenging; however, as explained Appendix \ref{section_SM_reps}, one way to reduce the number of extra supermassive fields
significantly is to abandon the requirement that the very massive extra matter
fill out complete $SU(5)$ multiplets.

\section{A Simple Model}\label{sec:simple}

\subsection{Basic scheme}\label{subsec:basic-scheme}

Before constructing models of calculable dynamical supersymmetry breaking that produce the pattern (\ref{simple}), 
we first realize a more modest goal and construct models in which the first and
second generations are composites of ${\it different}$ strongly coupled sectors.  If the first two 
generations arise from, say, dimension two operators in the high-energy theory and the third 
generation is elementary, the resulting Yukawa texture would be
\be
\label{hmm}
\left( \begin{array}{ccc}
\epsilon^2 & \epsilon\delta& \epsilon \\
\epsilon\delta &\delta^2 & \delta      \\
\epsilon & \delta    & 1 
\end{array}
\right)~,
\ee
with 
\be
\label{smallpar}
\epsilon = \Lambda_1 / M_{\rm flavor},~~\delta = \Lambda_2/ M_{\rm flavor}~.
\ee

While this is perhaps less elegant than obtaining the pattern (\ref{simple}), 
we will see that it is quite simple to realize in
practice.   One can therefore compare the relative complexity of the model building required to realize the
different textures and decide which seems more appealing.  In fact, as we will see, the simplest class
of models which realizes the texture (\ref{hmm}) can also, by variation of parameters, realize
the texture (\ref{simple}).  So it is quite natural to consider both patterns.

\subsection{Example}

\label{subsection_example}

The most obvious way to make a model with the pattern (\ref{hmm}) is to combine two 
of the calculable single-sector models
that produce a single composite generation which is dimension two in the UV theory, 
discussed in \S4.1\ of  \cite{Franco:2009wf}.

For instance, consider supersymmetric $SU(N_c)$ QCD with $N_c = 11$ and with $N_f = 12$ flavors of
quarks $Q, \tilde Q$, and a common quark mass $m \ll \Lambda$.     This theory has metastable vacua
which are evident in the weakly coupled magnetic dual description \cite{Seiberg}, valid at energies
$\ll \Lambda$.  The magnetic dual
is an $SU(N_f - N_c)$ gauge theory with $N_f$ flavors of magnetic quarks $q, \tilde q$, and a meson $\Phi$ which transforms
in the ${\bf({\rm Adj}+1)}$ of the $SU(12)$ flavor group but is a gauge singlet.
The magnetic superpotential is
\be
\label{Wmag}
W = h \tr(\Phi \tilde q q) - h \mu^2 \,\tr \Phi~,
\ee
where the second term arises due to the mass deformation of the electric theory.  Here, 
\be
\mu \sim \sqrt{m \Lambda}
\ee
and we can set $\Lambda_{\rm magnetic} = \Lambda$ (where the magnetic theory develops a Landau 
pole at $\Lambda_{\rm magnetic}$), so $h\sim 1$.

This theory breaks supersymmetry by the rank condition \cite{ISS}; the magnetic quarks develop a vacuum expectation value 
(vev) which breaks
the $SU(12)$ flavor symmetry to $SU(11)$, and $F_{\Phi} \neq 0$. We choose an embedding of the Standard
Model $SU(5)$ into the $SU(12)$ flavor group such that:
\be
\begin{array}{rcl}
Q & = & ({\bf 5}+ {\bf \overline{5}} + {\bf 1})+ {\bf 1}\\
\tilde{Q} & = & ({\bf \overline{5}} + {\bf 5} + {\bf 1})+ {\bf 1}
\end{array}
\label{embedding_12}
\ee
where the decomposition in parentheses indicates the embedding into $SU(11)$. The mesons of the magnetic theory can then be decomposed according to
\be
\Phi=\left(\begin{array}{ccc} Y_{1\times 1} & & Z^T_{1\times 11} \\
\tilde{Z}_{11\times 1} & & X_{11\times 11} \end{array}\right)~,
\ee
with $Y, Z, \tilde Z$ and $X$ transforming in the ${\bf 1}, {\bf \overline{11}}, {\bf 11},$ and ${\bf({\rm Adj}+1)}$ of
$SU(11)$.

In terms of $SU(5)$ quantum numbers, $X$ decomposes as
\be
X = ({\bf 10}+{\bf \bar{5}})+\left[2 \times {\bf 24} + {\bf 15} + {\bf \overline{15}}+{\bf \overline{10}}+ 2\times {\bf 5}+ {\bf \overline{5}}+3 \times {\bf 1}\right]~.
\label{Phi_0_decomposition}
\ee
We see that there is an entire Standard Model generation, and additional matter which can be given a large mass (at the
scale $\Lambda$) as in \cite{Franco:2009wf}, by adding appropriate ``spectators'' to the QCD dynamics and
deforming the superpotential by the mass term 
\be
\label{specmass}
W_{3} = \lambda \sum_{\bf R} \left( (Q \tilde Q)_{\bf R} S_{\bf {\overline R}} \right),
\ee
where the sum is over the representations in brackets in (\ref{Phi_0_decomposition}), except for the overall singlet $\tr\,X$ which breaks supersymmetry.  
Here $S_{\bf {\overline R}}$ are spectators added in the appropriate ${\it conjugate}$ representations.  After recalling
that the relation between the magnetic meson and $Q \tilde Q$ involves a power of $\Lambda$ to
canonically normalize the meson, the unwanted matter obtains masses of order
$\lambda \Lambda$ which can be a very high scale.  (We envision choosing $\Lambda$ just
below the GUT scale, for instance.)

The composite generation arising from $X$ is obviously of dimension two in the high-energy theory, and
therefore it will have Yukawa couplings suppressed by the ratio of scales $\Lambda/M_{\rm flavor}$.
The scalars in $X$ are pseudo-moduli which receive a calculable mass from loops in the magnetic
theory, of order $h^2 \mu/(4\pi)$.
Gauge mediation, with ``messengers'' coming both from the composite generation and some of the
additional components of $X$ and the magnetic quarks, will transmit masses of order $(g_{SM}^2/16\pi^2) \mu$ to the other Standard Model generations \cite{Franco:2009wf}.

It is now clear how to proceed to make a simple model which gives rise to the pattern of Yukawa couplings in
(\ref{hmm}), with two composite generations.  Consider an $SU(N_{c,1}) \times SU(N_{c,2})$ gauge theory
with $N_{f,1}$ flavors of quarks in the first gauge factor and $N_{f,2}$ in the second.  
If we choose $N_{c,i} = 11$, $N_{f,i}=12$, and independent
quark masses $m_i$ for the two sets of quarks, we end up with two copies of the previous model, with
supersymmetry-breaking scales $\mu_{1,2} = \sqrt{m_{1,2} \Lambda_{1,2}}$.
Gauge invariance forbids any additional marginal or relevant couplings in the electric theory, so in fact the most generic
renormalizable superpotential for the high-energy theory takes precisely the form we wish, though the
choice of parameters $m_{i} \ll \Lambda_i$ is only technically natural and would need to be retrofitted \cite{Dine:2006gm} in an 
acceptable construction.
Adding now an elementary pair of Higgs bosons and an elementary third generation, we will find precisely
the pattern of Yukawas in (\ref{hmm}), with $\epsilon$ and $\delta$ as in (\ref{smallpar}).

Problems from FCNCs in these type of models will be discussed in \S5 and appendix A.  
With the first and second generation sparticle masses $\sim $ 20 ${\rm TeV}$, only a moderate degeneracy among the two is 
required to avoid FCNCs. The soft masses of the first two generations come from the Coleman-Weinberg potential, 
generated after integrating out heavy fields, and are given by $\mu_1/4\pi$ and $\mu_2/4\pi$.  The $\mu_i$ should thus be
chosen to lie in the range $\sim 250 ~{\rm TeV}$ to avoid prohibitive FCNCs.\footnote{This introduces a new coincidence problem: 
why are the masses
generated by two unrelated sectors of strong dynamics relatively close to one another?  
We require $\mu_1$ and $\mu_2$ to be within roughly twenty percent of one another to avoid problems with FCNCs; the relevant 
constraints on similar
models will be discussed in great detail in \S5 and appendix A.  We note that
obtaining the two sectors from a single theory at higher energies, along the lines indicated in the next section, could
ameliorate this coincidence problem.}
 Gauge mediated
masses are dominated by the larger of these two scales.   There will be 8 additional messenger
pairs in the ${\bf 5 + \bf {\overline{5}}}$ of $SU(5)$, coming from the magnetic quarks and mesons in 
the two SQCD sectors.  Therefore, these models will have a Landau pole below the GUT scale.

In the discussion so far, we have not broken $R$-symmetry.  We can incorporate $R$-breaking by adding, for example, a further
superpotential deformation to the electric theory, $\Delta W_{el} \sim (Q \tilde Q)^2$.
This perturbation was studied in some detail in \cite{Essig:2008kz} (see also \cite{Koschade:2009qu}).  The
perturbation to the magnetic dual theory is
\be
\label{wfour}
W_{4} =\frac{1}{2} h^2 \mu_{\phi} \tr(\Phi^2)~.
\ee
This perturbation both explicitly breaks $R$-symmetry, and leads to a larger ${\it spontaneous}$ breaking, as
the $SU(11)$ singlet in $X$ develops a vev. 
After the addition of this coupling, the composite generation no longer arises strictly from $X$ --- instead, 
due to the mass terms from (\ref{specmass}) and (\ref{wfour}), each generation is now an admixture
of the ${\bf 10} + {\bf \overline{5}}$ from $X$ and one of the spectators.  However, for 
$\mu_\phi \ll \Lambda$, each generation is dominated by the composite field $X$, with the admixture from the spectator suppressed 
by the small parameter $\mu_{\phi}/\Lambda$.   To get interesting gaugino masses, $h^2 \mu_{\phi}$ should be chosen near 
the TeV scale, and if $\Lambda$ is near $M_{\rm GUT}$, the admixture is negligibly small.

\subsection{A landscape of simple models}\label{subsec:landscape}

We can derive the simple model in \S\ref{subsection_example} by starting with a high-energy theory consisting of a single $SU(N_c)$
gauge group with $N_f$ quark flavors together with an adjoint superfield $U$.  
The dynamics of this theory was studied in detail, in the presence of an adjoint superpotential, in 
\cite{Kutasov:1995ve, Kutasov:1995np, Kutasov:1995ss}.
Let us imagine that our theory
has a superpotential
\be
\label{adjsup}
W ~=~ \frac{g_{k+1}}{k+1}\,\Tr ( U^{k+1}) + \ldots+ g_1\, \Tr(U) ~=~\Tr(P_{k+1}(U))~,\ee
where $P_{k+1}(U)$ is a generic degree $k+1$ polynomial $P_{k+1} = \sum_{j=1}^{k+1} {\f{g_j}{j}} U^j$,
and $g_1$ should be interpreted as a Lagrange multiplier imposing the tracelessness constraint on
$U$.

The classical vacua of this theory can be found by 
setting the eigenvalues of the $N_c \times N_c$ traceless matrix $U$ equal to various roots of the equation
\be
\label{roots}
P^\prime(x) = \sum_{j=0}^{k} g_{j+1} x^j = \prod_{i=1}^{k} (x-a_i) ~ = ~0.
\ee
Let us assume that $P$ is sufficiently generic so that $a_i \neq a_j$ for $i \neq j$. In the vacuum where $N_i$ of the eigenvalues of $U$ are equal to $a_i$, and a total of $p$ different
$a_i$ appear as eigenvalues of $U$, 
the gauge group is broken as
\be
SU(N_c) \to \prod_{i=1}^{k} SU(N_i) \times U(1)^{p-1}
\ee
where $\sum_i N_i = N_c$.

The classical low-energy physics is that of a product of SQCD theories with $N_f$ quark flavors, but it is clear
that in the quantum theory the physics depends in detail on the precise values of the $a_i$, since
e.g. $a_i - a_j$ determines the masses of charged off-diagonal components of the $U$ field which serve
as bi-fundamentals connecting the different gauge factors.   As long as the $k$ roots $a_i$ in 
(\ref{roots}) are distinct, the adjoint
superfield gives rise to no massless excitations in any of these vacua.
Not all such partitions give rise to a theory with supersymmetric quantum vacua.  For instance, if any
of the $SU(N_i)$ factors has $N_i > N_f$, it suffers from a runaway to infinity in field space.  

Now, deforming the high-energy theory by a small quark mass $m$ for the $N_f$ quark flavors (small
compared to the effective adjoint mass in each vacuum), we obtain a
landscape of vacua with different $SU(N_i)$ gauge factors, each with $N_f$ quarks.  The different SQCD
sectors have different scales $\Lambda_i$, determined by matching scales at the value of the adjoint
mass. In particular, the scale of the $i$th theory is determined in terms of the scale $\Lambda$ of the original electric theory by 
\be
\Lambda_i^{3 N_i - N_f} = \Lambda^{2 N_c - N_f} g_{k+1}^{N_i} \prod_{j \neq i} (a_i - a_j)^{N_i - 2 N_j}.
\ee
This implies that the supersymmetry breaking scale of each $SU(N_i)$ theory is determined in terms of the scale of the parent $SU(N_c)$ 
gauge theory, the quark mass $m$, and the pattern of symmetry breaking encoded in (\ref{roots}). We can then anticipate generating a 
variety of vacua starting from one high-energy gauge theory, giving
rise to a discretuum of possible values of the parameters $\epsilon, \delta$ in \S \ref{subsec:basic-scheme}.  

Interestingly, in \S \ref{sec:adjoint}, we will also obtain models with the texture (\ref{simple}) from this class of gauge theories
with $k=2$.  So it is possible that one high-energy theory could give rise, in different vacua, to single-sector
models that each have a realistic phenomenology, with different explanations for the physics of flavor!

\section{SQCD with an adjoint}\label{sec:adjoint}

The previous section explored a class of models giving a Yukawa matrix (\ref{hmm}) based on 
two parameters $\epsilon$ and $\delta$. The rest of the paper is devoted to constructing calculable 
models with a ``dimensional hierarchy'', where the first and second generations arise from 
composite fields of dimension 3 and 2, respectively, while the third generation (denoted by $\Psi_3$) and the 
Higgs are elementary. Such models naturally give rise to the desired Yukawa texture (\ref{simple}) involving a 
single parameter $\epsilon.$

We now focus on the theory which appeared in \S \ref{subsec:landscape}:  the electric gauge theory will be
$SU(N_c)$ SQCD, with $N_f$ quarks $(Q_i, \t Q_j)$, and a field $U$ in the adjoint of the gauge group. While the analysis of \S 
\ref{subsec:landscape} was concerned with large adjoint masses (such that the adjoint could be integrated out in the low-energy 
theory), we will now be interested in the case where the adjoint mass is small and its dynamics remains important at low energies. This 
theory has been studied in detail in~\cite{Kutasov:1995ve, Kutasov:1995np, Kutasov:1995ss}, and 
we start by reviewing their conclusions.\footnote{See e.g.~\cite{Amariti} for a rather different construction 
of metastable vacua in SQCD with an adjoint.} 

\subsection{The electric theory}\label{subsec:adj-electric}

We begin by specializing to the case where
the adjoint has a general renormalizable superpotential
\begin{equation}\label{eq:adj-Wel1}
W_{el}= \frac{g_U}{3} \,\Tr U^3 + \frac{m_U}{2} \,\Tr U^2 + \lambda \, \Tr U\,.
\end{equation}
This superpotential does not have any metastable supersymmetry breaking vacua, which requires additional perturbations discussed 
below in \S \ref{subsec:adj-dsb}.   
Here `$\Tr$' means a trace over the gauge indices, while `$\tr$' will be used to indicate traces over flavor indices. 
$\lambda$ is a Lagrange multiplier field, imposing $\Tr U=0$. We denote the strong coupling scale by $\Lambda$. 
Calculability in the magnetic dual theory discussed below requires $m_U \ll \Lambda$. Higher dimensional 
operators $\Tr U^{k+1}$ with $k \ge 3$ are 
dangerously irrelevant and may influence IR physics if present~\cite{Kutasov:1995ss}. For now we focus on theories with $k = 2$, 
but we will have some discussion of theories with $k \ge 3$ in
\S \ref{subsec:exact}.

The matter content with its gauge and anomaly free global symmetry quantum numbers is (for $m_U=0$),
\begin{center}
\begin{tabular}{c|c|cccc}
&$SU(N_c)$&$SU(N_f)_L$&$SU(N_f)_R$&$U(1)_V$&$U(1)_R$\\
\hline
$Q$& $\Box$ & $\Box$ & $1$ & $1$ & $1-\frac{2}{3} \frac{N_c}{N_f}$ \nonumber\\
$\t Q$& $\overline \Box$ & $1$ & $\overline \Box$ & $-1$ & $1-\frac{2}{3} \frac{N_c}{N_f}$ \nonumber\\
$U$& Adj & $1$ & $1$ & $0$ & $\frac{2}{3}$ 
\end{tabular}
\end{center}
A nonzero mass $m_U$ breaks the $R$-symmetry. It will be useful to think of $m_U$ as a background superfield with $R$-charge $2/3$.

The superpotential has two critical points, $a_1$, $a_2$. The different classical vacua correspond to placing $r_1$ eigenvalues of $U$ equal to $a_1$, and $r_2=N_c-r_1$ eigenvalues equal to $a_2$. The gauge group is broken to
\begin{equation}\label{eq:break}
SU(N_c) \to SU(r_1) \times SU(r_2) \times U(1)\,.
\end{equation}
Imposing the tracelessness condition $r_1 a_1+ r_2 a_2=0$, the critical points are\footnote{Vacua with $r_1 = r_2$ can only exist for $m_U=0$. This case will not arise in our discussions.} 
\begin{equation}\label{eq:clvacua}
a_1 = \frac{r_2}{r_1-r_2}\,\frac{m_U}{g_U}\;,\;a_2 =- \frac{r_1}{r_1-r_2}\,\frac{m_U}{g_U}\,.
\end{equation}
The low energy theory splits into two decoupled SQCD sectors with only fundamental matter (as long as $m_U \neq 0$).
Quantum-mechanically, the vacua are stable if all $r_i \le N_f$; therefore, a necessary condition for the theory to have a stable vacuum is $N_f \ge N_c/2$. $N_f$ will also be restricted to $N_f < \frac{2}{3} N_c$ so that the magnetic theory is IR free. Summarizing, we will work in the range
\begin{equation}\label{eq:IRfree}
\frac{N_c}{2} < N_f < \frac{2}{3}N_c
\end{equation}
(The case $N_f=N_c/2$ is excluded because there are no magnetic quarks.)

An important role will be played by the two mesons
\begin{equation}\label{eq:our-def}
(M_1)_{ij}=\t Q_i Q_j\;\;,\;\;(M_2)_{ij}=\t Q_i U Q_j\,,
\end{equation}
where the gauge indices are contracted and suppressed. 
The moduli space is parametrized by these mesons and baryons 
(we refer the reader to~\cite{Kutasov:1995ss} for their definition, which will not be needed here), 
modulo classical relations. Notice that in~\cite{Kutasov:1995ss}, the dimension 3 meson was defined as
\begin{equation}\label{eq:KSS-def}
M^{KSS}_2=\t Q \left(U+ \frac{m_U}{2g_U}\right) Q\,.
\end{equation}
The redefinition $U \to U_s= U+\frac{m_U}{2g_U}$ amounts to setting $m_U=0$ and simplifies considerably the 
electric-magnetic duality discussion. However, we will work with the definition (\ref{eq:our-def}), where $M_2$ 
has classical scaling dimension 3, instead of being a linear combination of dimension 2 and dimension 3 fields. 
This simplifies the structure of the Yukawa couplings (\ref{hmm}) when we
later embed the first Standard Model generation inside $M_2$.

\subsection{The magnetic dual}\label{subsec:adj-magnetic}

The magnetic dual theory consists of SQCD, with gauge group 
$SU(\Ntilde_c=2N_f-N_c)$ and strong coupling scale $\t \Lambda$, 
$N_f$ quarks $(q,\tilde q)$, one magnetic adjoint field 
$\t U$, and two gauge singlet fields corresponding to 
the mesons (\ref{eq:our-def}). The theory has a superpotential\footnote{We are dropping a 
constant term which depends only on $g_U$. This becomes important when trying to match the 
gauge invariants $\Tr U^n \to \Tr \t U^m$. Also, (\ref{eq:adj-Wmag1}) differs slightly from the 
expression in~\cite{Kutasov:1995ss}; this is due to the meson definitions (\ref{eq:our-def}) and (\ref{eq:KSS-def}).}
\begin{eqnarray}\label{eq:adj-Wmag1}
W_{mag}&=&-\frac{g_U}{3}\,\Tr \t U^3+\frac{N_c}{2 \t N_c} m_U\,\Tr \t U^2 + \t \lambda \, \Tr \t U+ \nonumber\\
&+&\frac{g_U}{\hat \Lambda^2} \left(\frac{\t N_c-N_c}{2\t N_c}\frac{m_U}{g_U}\,\tr(M_1 q \t q)+ \tr(M_1  q \t U \t q)+ \tr(M_2 q \t q) \right)~.
\end{eqnarray}
The Lagrange multiplier $\t \lambda$ is introduced to impose $\Tr \t U =0$. The superpotential receives nonperturbative corrections~\cite{Kutasov:1995ss} that can be neglected near the origin of field space, where our metastable vacuum will be located.

The energy scale $\hat \Lambda$ appears because $M_1$ and $M_2$ are elementary, but have scaling dimensions 2 and 3, 
respectively. This dimensionful quantity is related to the electric ($\Lambda$) and magnetic ($\t \Lambda$) dynamical scales by
\begin{equation}
\Lambda^{2N_c-N_f} \t \Lambda^{2 \t N_c-N_f}= \left(\frac{\hat \Lambda}{g_U}\right)^{2N_f}\,.
\end{equation}
For $m_U=0$, the gauge and global (nonanomalous) symmetry transformations are
\begin{center}
\begin{tabular}{c|c|cccc}
&$SU(\t N_c)$&$SU(N_f)_L$&$SU(N_f)_R$&$U(1)_V$&$U(1)_R$\\
\hline
$q$& $\Box$ & $\overline \Box$ & $1$ & $\frac{N_c}{\t N_c}$ & $1-\frac{2}{3} \frac{\t N_c}{N_f}$ \nonumber\\
$\t q$& $\overline \Box$ & $1$ & $ \Box$ & $-\frac{N_c}{\t N_c}$ & $1-\frac{2}{3} \frac{\t N_c}{N_f}$ \nonumber\\
$\t U$& Adj & $1$ & $1$ & $0$ & $\frac{2}{3}$ \nonumber\\
$M_1$ & $1$& $\Box$ & $\overline \Box$& $0$&$2-\frac{4}{3} \frac{N_c}{N_f}$\nonumber\\ 
$M_2$ & $1$& $\Box$ & $\overline \Box$& $0$&$\frac{8}{3}-\frac{4}{3} \frac{N_c}{N_f}$\nonumber\\ 
\end{tabular}
\end{center}
Notice the different $R$-charge of $M_1$ and $M_2$ (which can be read off directly in the electric theory). A nonzero mass $m_U$ breaks the $R$-symmetry.

In the range (\ref{eq:IRfree}), the magnetic theory is IR free and the K\"ahler potential can be expanded
\begin{equation}\label{eq:adj-K}
K= \frac{1}{\alpha_1 |\Lambda|^2}\,\tr(M_1^\dag M_1)+\frac{1}{\alpha_2 |\Lambda|^4}\,\tr(M_2^\dag M_2)+\frac{1}{\alpha_3 }\,\tr(q^\dag q + \t q \t q^\dag)+\frac{1}{\alpha_4}\,\Tr(\t U^\dag \t U)+ \ldots
\end{equation}
where $\alpha_i$ are order one positive numbers and `$\ldots$' include interaction terms. The canonically normalized mesons are
\begin{equation}\label{eq:adj-mesons}
\Phi:=\frac{M_1}{\sqrt{\alpha_1}\,\Lambda}= \frac{\t Q Q}{\sqrt{\alpha_1}\,\Lambda}\;\;,\;\;\Phi_U:=\frac{M_2}{\sqrt{\alpha_2}\,\Lambda^2}=\frac{\t Q U Q}{\sqrt{\alpha_2}\,\Lambda^2}\,.
\end{equation}
Similarly, replacing $q \to \sqrt{\alpha_3} q$, $\t q \to \sqrt{\alpha_3} \t q$ and $\t U \to \sqrt{\alpha_4} \t U$ gives canonical kinetic terms to the adjoint and magnetic quarks. Henceforth, only canonically normalized fields will be used.

The superpotential then becomes
\begin{eqnarray}\label{eq:adj-Wmag2}
W_{mag}&=&\frac{\t g_U}{3}\,\Tr \t U^3+\frac{\t m_U}{2} \,\Tr \t U^2 + \t \lambda' \, \Tr \t U+ \nonumber\\
&+&\frac{h}{\Lambda}\,\left[c_1 \t m_U \,\tr (\Phi q  \t q) + c_2\,\tr(\Phi q \t U \t q) \right]+ h\,\tr(\Phi_U q \t q)\,.
\end{eqnarray}
The parameters introduced here are related to the previous ones by
\begin{eqnarray}\label{eq:adj-params}
\t g_U&:=& - (\alpha_4)^{3/2} g_U\;,\;\t m_U:= \frac{\alpha_4 N_c}{\t N_c} m_U\;,\;h:=\sqrt{\alpha_2} \alpha_3\,\frac{g_U \Lambda^2}{ \hat \Lambda^2}\nonumber\\
c_1&:=&\left(\frac{\alpha_1 \alpha_4}{\alpha_2} \right)^{1/2} \frac{N_c-N_f}{\t g_UN_c}\;,\;c_2:= \left(\frac{\alpha_1 \alpha_4}{\alpha_2} \right)^{1/2}\;,\; \t \lambda':=\sqrt{\alpha_4} \,\t \lambda\,.
\end{eqnarray}
Also, $\t m_U \ll \Lambda$ is required for calculability in the magnetic theory (although in the opposite limit, $\t m_U \gg \Lambda,$ the adjoint may be integrated out of the electric theory to produce the models of \S \ref{subsec:landscape}).

We end this analysis by pointing out the following interesting consequence of the duality. All the interactions between the meson $\Phi$ and the rest of the fields of the magnetic theory are suppressed by $1/\Lambda$. At energies $E \ll \Lambda$, $\Phi$ approximately decouples from the rest of the system. In particular, while the trilinear coupling between $\Phi_U$ and the magnetic quarks is order $h$, the corresponding interaction for $\Phi$ is only order $h \t m_U / \Lambda$. This difference can be understood as follows: When $\t m_U=0$ the $U(1)_R$ symmetry presented before forbids a coupling $\Phi q \t q$. Turning on a nonzero mass and treating it as a spurion superfield, the only trilinear coupling allowed by $R$-symmetry is $(\t m_U/\Lambda) \Phi q \t q$.

\subsection{Metastable supersymmetry breaking}\label{subsec:adj-dsb}

The low energy theory (\ref{eq:adj-Wmag2}) contains a massive adjoint $\t U$, magnetic 
quarks $(q, \t q)$ interacting with a meson $\Phi_U$, and an extra meson $\Phi$ whose 
interactions with the other fields are suppressed by $1/\Lambda$. The $(\Phi_U, q, \t q)$ 
sector is very similar to the magnetic theory studied by Intriligator, Seiberg, and Shih (ISS) 
in \cite{ISS}, although the corresponding electric theories are quite different. For instance 
$\Phi_U$ is of dimension 3 in the UV, while the ISS meson has scaling dimension 2.

We focus on vacua with $\langle \Tr \,\t U^2 \rangle = 0$, corresponding to 
$r_1=N_f$, $r_2=N_c-N_f$ in (\ref{eq:clvacua}).  For this choice of 
parameters the magnetic gauge group is unbroken. In general, it will also be convenient to set $\t N_c=1$, to reduce the amount of additional matter (see \S\ref{sec:adj-ssector}). Then the magnetic gauge group is trivial, there is no magnetic adjoint, and the magnetic superpotential simplifies to
\begin{equation}\label{eq:adj-Wmag3}
W_{mag}=c_1 h \frac{\t m_U}{\Lambda}\,\tr(\Phi q \t q)+ h\,\tr(\Phi_U q \t q)\,.
\end{equation}
Importantly for the low energy physics, in this case there is an additional $R$-symmetry under which the mesons have charge 2, while the magnetic quarks have charge 0. This symmetry, which is anomalous, will be denoted by $U(1)_R'$. Once the Standard Model gauge group is embedded in the symmetry group of the theory, we will need to break $U(1)_R'$ in order to generate large enough gaugino masses.

In the low-energy theory, $\t m_U/\Lambda$ appears as a free parameter which determines how strongly the meson $\Phi$ 
couples to the magnetic quarks. For pedagogical purposes, we first restrict ourselves to the limit $\t m_U \ll \Lambda$, which simplifies the 
analysis considerably. While this limit can lead, for a careful choice of parameters, to a phenomenologically viable model 
that is not in conflict with current limits from FCNCs (see \S\ref{subsec:approx.}), 
larger values of $\t m_U$  (\S\ref{subsec:decoupling}) or additional superpotential interactions (\S \ref{subsec:exact}) are desirable.

In this weakly coupled description, a supersymmetry breaking vacuum is generated once a term $\tr\, \Phi_U$ is added to the superpotential.\footnote{We break supersymmetry predominantly with $\Phi_U$ because the interactions of $\Phi$ with the magnetic quarks are suppressed by $\t m_U / \Lambda \ll 1$. Other deformations are explored below.} Following~\cite{Essig:2008kz, Giveon:2007ef}, the $U(1)'_R$ symmetry will be broken by adding a small explicit breaking term proportional to $\tr \,\Phi_U^2$. Furthermore, in order to avoid an exactly massless superfield, a mass term $\tr \, \Phi^2$ is needed.

Summarizing, the superpotential including the minimal set of deformations required to construct a realistic model of supersymmetry breaking is
\begin{equation}\label{eq:adj-Wmag4}
W_{mag}= c_1 h \frac{\t m_U}{\Lambda}\,\tr(\Phi q \t q)+ \frac{1}{2}m_\Phi\,\tr\,\Phi^2 + \left[- h \mu^2\,\tr\,\Phi_U+ h\,\tr(\Phi_U q \t q)+ \frac{1}{2}h^2\muphi\,\tr(\Phi_U^2)\right]\,.
\end{equation}
To facilitate the interpretation of the model, the fields and interactions that will be responsible of breaking supersymmetry have been collected inside square brackets. The deformation parameters $m_\Phi$, $\mu$ and $\muphi$ should be parametrically smaller than the dynamical scale $\Lambda$ so that microscopic corrections to the K\"ahler potential can be neglected.

Equation (\ref{eq:adj-Wmag4}) is the full superpotential when $\t N_c = 1$. For $\t N_c > 1$, it is straightforward to add the adjoint and interactions described in (\ref{eq:adj-Wmag2}); in this case, the formulas below are still valid in the vacuum $\langle \Tr \t U^2 \rangle = 0$.

Foreseeing the use of this theory as a single-sector model of supersymmetry breaking, 
we point out that certain off-diagonal components of $\Phi_U$ and $\Phi$ will 
be identified with the first and second Standard Model generations. Of course, such components 
cannot have large vector-like supersymmetric masses via superpotential terms 
(\ref{eq:adj-Wmag4}) that couple them to conjugate fields. The Standard Model composite 
generations will be made massless by introducing heavy spectator fields 
coupled to the unwanted conjugate fields. However, for now we will analyze 
the theory with superpotential (\ref{eq:adj-Wmag4}) and no extra fields.

In the electric theory, the deformations added to (\ref{eq:adj-Wmag3}) to arrive at (\ref{eq:adj-Wmag4}) 
correspond to perturbing (\ref{eq:adj-Wel1}) by
\begin{equation}\label{eq:DWel}
\Delta W_{el} \sim \lambda_Q \tr(QU\t Q)+\frac{\lambda_1}{\Lambda_0} \tr (Q \t Q)^2+ \frac{\lambda_2}{\Lambda_0^3} \tr (Q U \t Q)^2
\end{equation}
where $\Lambda_0$ is some UV scale satisfying $\Lambda_0 \gg \Lambda$. In particular, the 
Yukawa interaction $\lambda_Q \tr (Q U \t Q)$ in (\ref{eq:DWel}) gives rise to the supersymmetry-breaking source term  
$- h \mu^2 \tr (\Phi_U)$ appearing in (\ref{eq:adj-Wmag4}). Thus $\mu$ is related to the parameters of the electric theory by 
\begin{eqnarray}
h \mu^2 := \lambda_Q \sqrt{\alpha_2} \Lambda^2 \ , \ \mu := \sqrt{\frac{\lambda_Q}{\alpha_3 g_U}} \hat \Lambda,
\end{eqnarray}
The parametric separation of scales $\mu \ll \Lambda$ required for calculability and metastability in the magnetic theory arises from the smallness of the dimensionless coupling $\lambda_Q$, as contrasted with the dimensionful quark mass $m$ of \cite{ISS}. Indeed, all the deformations introduced in (\ref{eq:DWel}) arise from marginal and irrelevant interactions in the electric theory. More general perturbations will be discussed momentarily.

Since $\muphi$ comes from an irrelevant operator in the electric theory, we naturally have $\muphi \ll \mu$. 
The analysis then proceeds as in~\cite{Essig:2008kz}. In the limit $\muphi \to 0$ supersymmetry is broken at 
tree level by the rank condition, and $\Phi_U$ is stabilized at the origin due to one-loop effects. For finite 
$\muphi \ll \mu$, the $U(1)'_R$ is explicitly broken and supersymmetric vacua appear at a distance 
$\mu^2/\muphi$ from the origin.  At tree-level, there are no supersymmetry breaking vacua.  
However, supersymmetry can be broken in a long-lived metastable vacuum that lies close to the origin 
when one-loop quantum corrections 
are included (see below) \cite{Essig:2007xk,Essig:2007vu}. 
The tunneling from the metastable vacuum to the supersymmetric vacua is highly suppressed for $\muphi\ll\mu$.
Of course, there are also supersymmetric vacua at large values of $\Phi_U$, whose existence crucially relies on 
(calculable) non-perturbative effects 
\cite{Affleck:1983mk}, but as in \cite{ISS, Amariti} the longevity of the metastable vacuum here is 
guaranteed by the hierarchy $\mu / \Lambda \ll 1$. Finally, the theory possesses a large number of additional 
vacua labeled by the possible partitions (\ref{eq:break}) of the gauge group; stability of the vacuum 
with $\langle \Tr \,\t U^2 \rangle = 0$ against potential transitions into such vacua may be guaranteed 
provided  $\mu \ll \tilde m_U$, which is readily accommodated.

Let us now analyze the pattern of supersymmetry breaking in more detail.  We parameterize the fields as
\begin{equation}\label{eq:paramPhi}
\Phi_U= \left(\begin{matrix} Y_{U,\,\tilde N_c \times \tilde N_c} & Z^T_{U,\,\tilde N_c \times (N_f-\t N_c)} \\ \tilde Z_{U,\,(N_f-\t N_c) \times \tilde N_c} &X_{U,\,(N_f-\t N_c) \times (N_f-\t N_c)}\end{matrix} \right)\;,\;\Phi= \left(\begin{matrix} Y_{\tilde N_c \times \tilde N_c} & Z^T_{\tilde N_c \times (N_f-\t N_c)} \\ \tilde Z_{(N_f-\t N_c) \times \tilde N_c} &X_{(N_f-\t N_c) \times (N_f-\t N_c)}\end{matrix} \right)\;,
\end{equation}
\begin{equation}\label{eq:paramq}
q^T=\left( \begin{matrix} \chi_{\tilde N_c \times \tilde N_c} \\ \rho_{(N_f-\t N_c) \times \tilde N_c} \end{matrix}\right)\;,\;\tilde q=\left( \begin{matrix} \tilde \chi_{\tilde N_c \times \tilde N_c} \\ \tilde \rho_{(N_f-\t N_c) \times \tilde N_c} \end{matrix}\right)\,.
\end{equation}
We will not present the spectrum of this model in detail\footnote{We refer the reader to~\cite{Essig:2008kz}.}, 
but only focus on the fields $\rho$, $\rhot$, $\ZU$, $\ZUt$, $Z$, and $\Zt$.  
Integrating out these fields generates the (bosonic) Coleman-Weinberg potential, which in general is given by~\cite{Coleman:1973jx}
\begin{equation}\label{eq:Vcw1}
V_{CW}=\frac{1}{64 \pi^2}\,{\rm STr}\,M^4\,{\rm log}\,\frac{M^2}{\Lambda_{\rm cut}^2}\,,
\end{equation}
where $M$ is the mass matrix of the fields being integrated out and $\Lambda_{\rm cut}$ is some high-energy cut-off.  
The superpotential for the fields that generate the Coleman-Weinberg potential that will lift the tree-level runaway direction $X_U$ is 
\begin{equation}\label{eq:Wmassive}
W \supset h \left( \begin{matrix} \rho& Z_U & Z\end{matrix} \right) 
\left( \begin{matrix}  X_U & \chi \;\;& \f{c_1 \t m_{U}}{\Lambda} \chi \\ 
 \t \chi  & h\muphi & 0 \\
\f{c_1 \t m_{U}}{\Lambda} \t \chi \;\; & 0 & m_\Phi 
\end{matrix} \right) \left( \begin{matrix} \tilde \rho\\ \ZUt \\ \t Z\end{matrix} \right)\,,
\end{equation}
where $\chi\chit$ is given by (\ref{eq:XAvev}).  
Since we take $\t m_U/\Lambda \ll 1$, the $Z$, $\Zt$ fields completely decouple from the $\rho, \rhot, Z_U, \ZUt$ sector.  
Moreover, the supersymmetry breaking field $X_U$ couples in this limit only to the $\rho, \rhot, Z_U, \ZUt$ sector, and we 
can focus on the fermion mass matrix
\begin{equation}\label{eq:reducedFermion}
M_f = h 
\left( \begin{matrix}  X_U & \chi \\ 
 \t \chi  & h\muphi 
\end{matrix} \right).
\end{equation}
The bosonic components of $\rho, \rhot, Z_U, \ZUt$ will have masses given by 
\begin{equation}\label{eq:bosons}
M_b =  
\left( \begin{matrix}  M_f^\dagger M_f & -h^* F_{X_U}^* \\
-h F_{X_U} & M_f M_f^\dagger 
\end{matrix} \right), \;{\textrm{with }}\; 
-F_{X_U}^* = h  
\left( \begin{matrix}  -\mu^2 + h \muphi X_U& 0 \\
0 & 0
\end{matrix} \right).
\end{equation}

The analysis proceeds now exactly as in \cite{Essig:2008kz}, and we may borrow the results from there. 
Near the origin of field space, the Coleman-Weinberg potential from integrating out $\rho, \rhot, Z_U$, and $\ZUt$ is
\begin{equation}\label{eq:adj-CW}
V_{CW}= m_{CW}^2 |X_U|^2+ \ldots
\end{equation}
where `$\ldots$' refers to higher order interactions and mixings with $X$ that can be neglected. The ``Coleman-Weinberg mass'' is
\begin{equation}\label{eq:mcw}
m_{CW}^2= b |h^2 \mu|^2\;\;,\;\;b=\frac{\log \,4 -1}{8 \pi^2} \t N_c\,.
\end{equation}
Combining (\ref{eq:adj-CW}) with the tree-level potential computed from (\ref{eq:adj-Wmag4}), 
\be
V_{\rm tree} = (N_f-\Nct) |-h \mu^2 + h^2 \muphi X_U|^2,
\ee
we find
\begin{equation}\label{eq:XAvev}
\langle h X_U \rangle \approx\,\frac{\mu^2 \muphi^*}{b|\mu|^2+ |\muphi|^2} \approx \frac{\muphi^*}{b}\;\;,\;\;\langle \chi \t \chi \rangle \approx \mu^2
\end{equation}
and
\begin{equation}
|W_{X_U}|\approx |h \mu^2|\,.
\end{equation}
Importantly for the low energy phenomenology, the vev of $X_U$ is larger 
than $\muphi^*$ by the inverse loop factor $1/b\sim 16\pi^2$. Hence, the spontaneous breaking of the $R$-symmetry 
is parametrically larger than the explicit one, and gaugino masses can be sufficiently large. 
Corrections suppressed by $1/\Lambda$ have been neglected. 

The field $\Phi$ is stabilized supersymmetrically, 
\begin{equation}
W_\Phi = 0\;,\;\langle X \rangle = 0\;,\;\langle Y \rangle \approx -c_1 \frac{\t m_U}{\Lambda}\,\frac{h \mu^2}{m_\Phi}\,,
\end{equation}
where we have neglected corrections of $\OO(\muphi\mu^2 \t m_{U}^3/(m_\Phi^2\Lambda^3))$.
From the $F$-term for the magnetic quarks, we find
\begin{equation}
\langle Y_U \rangle = - c_1 \frac{\t m_U}{\Lambda}\,\langle Y \rangle \,.
\end{equation}
The rest of the fields are stabilized at the origin. The hierarchy $\muphi \ll \mu \ll \Lambda$ ensures that the vacuum is parametrically long-lived against transitions into the various supersymmetric vacua~\cite{Essig:2008kz}.
The theory receives microscopic corrections controlled by $\t m_U / \Lambda$ and $\mu/\Lambda$, which are parametrically suppressed
 compared to the IR effects we have discussed. At this order, it is consistent to set $\langle Y \rangle = \langle Y_U \rangle = 0$.  
Moreover, (\ref{eq:adj-Wmag4}) implies that there are one-loop contributions mixing $X$ and $X_U$,
\be
V_{1-loop} \sim m_{CW}^2\,{\rm Re}\left(\frac{\t m_U}{\Lambda}X^* X_U\right)\,.
\ee
This is negligible in the limit $\t m_U \ll \Lambda$.  
Finally, we note that the unbroken global symmetry is 
\be
SU(N_f - \t N_c) \times U(1)\,.
\ee 
In \S \ref{sec:adj-ssector}, we will weakly gauge and identify a subgroup of $SU(N_f-\Nct)$ with the Standard Model 
gauge group.  This will mean that part of the $X_U$, $X$, $Z_U$, $\ZUt$, $\rho$, $\rhot$, $Z$, and $\Zt$ will have 
Standard Model gauge charges.  In particular, we will identify part of $X_U$ and $X$ with the first and second generation 
Standard Model fermions.   

\subsection{More general superpotential perturbations}\label{subsec:more-general}

Let us summarize what we have done so far:
\begin{enumerate}
\item We have constructed a metastable vacuum based on the (almost decoupled) 
sector $(\Phi_U, q, \t q)$, by having superpotential terms that are linear and quadratic in $\Phi_U$; see (\ref{eq:adj-Wmag4}).
\item The extra meson $\Phi$ has been lifted by adding an appropriate mass term, 
which is naturally large in the magnetic theory once $U(1)_R'$ is broken. This sector 
is decoupled from the supersymmetry breaking sector at leading order in $\t m_U/\Lambda$.
Later on, one chiral generation from this sector will be re-coupled.
\item In the metastable vacuum, the magnetic gauge group is completely Higgsed at 
the scale $\langle \chi \t \chi \rangle = \mu^2$. The magnetic adjoint $\t U$ is massive 
and its interactions with the rest of the fields are suppressed by $1/\mu$ and $1/\Lambda$. 
Or, in the case of $\t N_c=1$, the magnetic theory has no adjoint to begin with, as explained around (\ref{eq:adj-Wmag4}).
\end{enumerate}

In the high-energy electric gauge theory, we have allowed only specific  
marginal and irrelevant operators (\ref{eq:DWel}). The aim of this 
subsection is to discuss what happens when more general deformations are allowed.

Adding a $U^4$ piece changes the chiral ring and 
introduces extra degrees of freedom in the low energy theory. The resulting low energy phenomenology will be analyzed in \S \ref{subsec:exact}. On the other hand, adding $U^n$ factors (with $n \le 3$) to any superpotential term 
containing the mesons $(Q \t Q)$ and/or $(Q U \t Q)$, modifies negligibly the low energy theory. This is because we are considering a 
vacuum where the magnetic adjoint does not have a vev, and it has suppressed couplings to the supersymmetry breaking sector.

We are thus left with the possibility of adding irrelevant operators up to dimension 6, formed from the two mesons. 
One possibly dangerous 
term, which may give large FCNCs, arises from the dimension 5 operator $(\t Q Q) (\t Q U Q)$ --- 
this results in a mixing between $\Phi$ and $\Phi_U$ in the low energy 
magnetic theory. 
The full magnetic superpotential arising from marginal and irrelevant deformations of the electric superpotential, up to dimension 6, is of the 
form
\begin{eqnarray}
W_{mag}&=& -h\mu^2\,\tr\,\Phi_U+\frac{1}{2} m_\Phi \,\tr\, \Phi^2+ \Delta m\,\tr\,\Phi \Phi_U + \alpha\,\tr \,\Phi^3+\nonumber\\
&+& \frac{1}{2}h^2 \muphi\,\tr\,\Phi_U^2+c_1 h \frac{\t m_U}{\Lambda}\,\tr(\Phi q \t q)+ h\,\tr(\Phi_U q \t q)\,.
\end{eqnarray}
The cubic term does not alter our analysis of the metastable vacuum near the origin of field space. Furthermore, as long as $(\Delta m)^2 \lesssim m_\Phi \muphi$, the results of the previous subsection are approximately correct. 

However, for $(\Delta m)^2> m_\Phi \muphi$, the computation of the metastable vacuum receives important corrections. In this range there 
is still a metastable vacuum, but now both $\Phi_U$ and $\Phi$ play a role in the supersymmetry breaking dynamics, and 
their scalar components (part of which will become the first and second generation sfermions) 
receive direct supersymmetry breaking masses from the Coleman-Weinberg potential. This alternative will be explored, and exploited, in \S \ref{subsec:decoupling}.

\section{Single-sector supersymmetry breaking}\label{sec:adj-ssector}

The model of \S \ref{sec:adjoint} with magnetic superpotential 
(\ref{eq:adj-Wmag4}) will now be used to construct a ``single-sector'' supersymmetry breaking model in which 
some Standard Model generations are composite mesons of the strongly coupled electric theory.  
In \S \ref{subsec:fermions}, we discuss a simple embedding of the first and second generation 
Standard Model fermions into the mesons of the supersymmetry breaking sector.  We show how this generates the desired 
fermion Yukawa matrix, (\ref{simple}), and thus naturally produces the observed flavor hierarchy.  
In \S \ref{subsec:spectrum}, we estimate the parametric contributions to various sparticle masses.  While the gaugino masses are generated from 
gauge mediation only, the sfermions may obtain a mass from gauge mediation or directly from the 
supersymmetry breaking sector (in particular, from the one-loop Coleman-Weinberg potential).  

Constraints on the sfermion masses from flavor-changing neutral currents (FCNCs) are discussed in \S \ref{subsec:comments1}.  
Although the sfermion masses are diagonal in the flavor basis in which the fermion Yukawa matrices 
take on the texture of (\ref{simple}), large off-diagonal sfermion mass terms may be generated after diagonalizing 
the fermion Yukawas.  This can lead to large FCNCs unless the sfermion masses of first two generations are  
roughly the same (universal) or are both very heavy (decoupled). Successful model-building then amounts to finding various limits of the adjoint model that give rise to soft terms compatible with FCNC and other constraints. We will reserve a discussion of specific parametric limits and viable soft spectra for \S \ref{sec:sols}.

\subsection{MSSM generations from composites}\label{subsec:fermions}

A simple single-sector supersymmetry breaking model can be constructed by embedding the first Standard Model generation 
inside the meson $\Phi_U$ and the second generation inside the meson $\Phi$ 
(the embeddings are described in detail below).  
The third generation will come from an additional elementary field, which we denote 
by $\Psi_3$.  The fields $\Phi$ and $\Phi_U$ were defined in (\ref{eq:adj-mesons}) but 
are reproduced here schematically for convenience:
\be
\Phi_U \sim \frac{\t Q U Q}{\Lambda^2}\;\;,\;\;\;\Phi \sim \frac{\t Q Q}{\Lambda}\,.
\ee
While both $\Phi_U$ and $\Phi$ are dimension one fields at low energies in the magnetic theory, 
they are dimension three and two fields, respectively, in the UV electric theory.  

The fermion Yukawa couplings will be generated at a ``flavor scale'' $M_{flavor}$, where 
the electric theory is weakly coupled, through couplings between the Standard Model fields 
contained inside $\Qt U Q$, $\Qt Q$, and $\Psi_3$ and an elementary Higgs field, $H$,
\be\label{eq:Yuk}
W_{Yuk} & \supset & \frac{1}{M_{\rm flavor}^4} (QU\t Q) H (QU\t Q) + \frac{1}{M_{\rm flavor}^3} (Q\t Q) H (QU \t Q) + \f{1}{M_{\rm flavor}^2} (Q\t Q) H (Q\t Q)+\nonumber \\
& &  + \f{1}{M_{\rm flavor}} (Q\t Q) H \Psi_3 + \f{1}{M_{\rm flavor}^2} (Q\t U Q) H \Psi_3 +\Psi_3 H \Psi_3.
\ee
We have neglected $\OO(1)$ dimensionless couplings. Since $\Qt U Q$, $\Qt Q$ and $\Psi_3$ are dimension three, two, and, one, respectively, the generated 
Yukawa couplings are suppressed by different powers of the flavor scale $M_{\rm flavor}$.  

At low energies, this Yukawa superpotential becomes
\begin{eqnarray}
W_{Yuk} &\supset&  \frac{\Lambda^4}{M_{\rm flavor}^4} \Phi_U H \Phi_U + \frac{\Lambda^3}{M_{\rm flavor}^3} \Phi H \Phi_U +
 \f{\Lambda^2}{M_{\rm flavor}^2} \Phi H \Phi +\nonumber\\
&+& \f{\Lambda}{M_{\rm flavor}} \Phi H \Psi_3 + \f{\Lambda^2}{M_{\rm flavor}^2} \Phi_U H \Psi_3 +\Psi_3 H \Psi_3. 
\end{eqnarray}
Setting $\epsilon = \Lambda/M_{\rm flavor}$ gives the following fermion Yukawa matrix (up to $\OO(1)$ dimensionless couplings)
\be
\left( \begin{array}{ccc}
\epsilon^4 & \epsilon^3 & \epsilon^2 \\
\epsilon^3 &\epsilon^2 & \epsilon      \\
\epsilon^2 & \epsilon    & 1 
\end{array}
\right),
\ee
which will generate the desired flavor hierarchy for $\epsilon\sim 10^{-1}$.  
Note that it requires $\Lambda \sim 10^{-1} M_{\rm flavor}$, so that the strong coupling scale of the electric theory 
cannot be too much below the ``flavor'' scale.  

We now describe the embedding of the Standard Model fields inside the supersymmetry breaking mesons in more detail. 
To present our results in a compact way, an $SU(5)$ GUT notation will be adopted, but the Standard Model gauge group 
$SU(3)_C \times SU(2)_L \times U(1)_Y$ can be easily used instead. 
The latter embedding will be explored in Appendix \ref{section_SM_reps} and has the advantage that it generates
less additional heavy Standard Model charged matter that change the RG running of the Standard Model 
gauge couplings  --- in particular, Landau poles 
(which we discuss in Appendix \ref{sec:unification}) can be pushed to much higher energy scales. 

The minimal choice for the number of flavors and colors of the electric theory corresponds to
$$
N_f= 12\;,\;\t N_c =1 \;\Rightarrow\; N_c=23
$$
The $SU(N_f=12)$ global symmetry is broken to $SU(N_f-\Nct=11)$ by the vacuum expectation value $\chi \t \chi = \mu^2$ (see (\ref{eq:XAvev})). 
The Standard Model GUT group is a weakly gauged $SU(5)$ subgroup of $SU(11)$, with the following embedding of 
$SU(5)$ into $SU(12)$:
\be\label{eq:embed}
Q \, \sim \, ({\bf 5} + {\bf \bar{5}} + {\bf 1}) + {\bf 1}\;,
\t Q \, \sim \, ({\bf \bar{5}} + {\bf 5} + {\bf 1}) + {\bf 1},
\ee
where the representations in round brackets denote the embedding into $SU(11)$.  

The mesons of the magnetic theory decompose as (see (\ref{eq:paramPhi}))
\begin{equation}\label{decomposition_PhiU_Phi}
\Phi_U= \left(\begin{matrix} Y_{U,\,1 \times 1} & Z^T_{U,\,1 \times 11} \\ \tilde Z_{U,\,11 \times 1} &X_{U,\,11 \times 11}\end{matrix} \right)\;,\;\Phi= \left(\begin{matrix} Y_{1 \times 1} & Z^T_{1 \times 11} \\ \tilde Z_{11 \times 1} &X_{11\times 11}\end{matrix} \right)\;,
\end{equation}
The fields $(Y_i, \chi, \t \chi)$ fields are all singlets under the Standard Model gauge group, while 
$X_U$ and $X$ decompose as
\begin{equation}\label{eq:reps2}
({\bf 10}+ \bar{\bf 5}) + \left[2 \times {\bf 24} + {\bf 15} + {\bf \overline{15}} + {\bf \overline{10}} + 2 \times {\bf 5} + {\bf \bar 5} + 3\times {\bf 1} 
\right],
\end{equation}
where the representations in round brackets will form the desired Standard Model fermions and 
the matter in square brackets represents additional matter that we will want to remove.  

The unwanted matter can be removed by the addition of spectator fields $S_{\bf \bar{R}}$ for each 
representation ${\bf R}$ in square brackets 
(except the singlet piece $\Tr(X_U)$, which participates in supersymmetry breaking)
and with superpotential couplings 
\bea\label{eq:additionalMatter}
W_{el} & \supset & \lambda_{1{\bf R}} \sum_{\bf{\bar{R}}} S_{1 {\bf{\bar{R}}}} (Q \t Q)_{\bf R} +  
\lambda_{2{\bf R}} \f{1}{\Lambda_0} \sum_{\bf{\bar{R}}} S_{2 {\bf{\bar{R}}}} (Q U \t Q)_{\bf R} \nonumber\\
\to W_{mag} & \supset & 
\lambda_{1{\bf R}} \Lambda \sum_{\bf{\bar{R}}} S_{1 {\bf{\bar{R}}}} X_{\bf R} +  
\lambda_{2{\bf R}} \f{\Lambda^2}{\Lambda_0} \sum_{\bf{\bar{R}}} S_{2 {\bf{\bar{R}}}} X_{U,{\bf R}}.
\eea
The unwanted matter will now have masses of order $\Lambda$ and $\Lambda^2/\Lambda_0$, where 
$\Lambda_0$ is some UV scale above $\Lambda$.   

We also include spectators that pair up with $Z$ and $\tilde Z$, which are also charged under the Standard Model 
gauge group. 
It is worth briefly explaining why we can include spectators to remove the unwanted $Z,  \tilde Z$ particles
in this model, but not e.g.~in the models of \cite{Franco:2009wf}.  In ISS-like models, the $Z$ and $\Zt$ are in the
same multiplet as the magnetic meson that breaks supersymmetry by the rank condition, and they receive a tree-level
supersymmetry-breaking mass.  This is because they mix with the $\rho$ components of the magnetic quarks, which
obtain a mass from the $\tilde q \Phi q$ coupling in the magnetic superpotential. Therefore,
they play an important role in the calculation of the one-loop Coleman-Weinberg potential, and altering the spectrum
of $Z, \Zt$-mesons, even if it could be done without creating instabilities, would drastically affect the model.  
In this model, in contrast, there are two magnetic mesons, and only $\Phi_U$ is playing a role in the
supersymmetry breaking, while $\Phi$ is almost a spectator to the dynamics.  Therefore, the $Z, \tilde Z$
mesons play no role in the Coleman-Weinberg computations, and can be safely given a large mass of order $\Lambda^2 / \Lambda_0$ from the coupling (\ref{eq:DWel}), or an even larger mass of order
$\lambda \Lambda$ by adding appropriate spectators.  

Once the chiral deformation (\ref{eq:additionalMatter}) is turned on, the $({\bf 10}+ \bar{\bf 5})$ Standard Model fermions from 
$X$ and $X_U$ (see (\ref{decomposition_PhiU_Phi})) acquire masses only from the superpotential coupling (\ref{eq:Yuk}). More 
precisely, due to the $\muphi$ perturbation the chiral fermions have a very small admixture with the spectators. 
This mixing is of order $(\muphi \Lambda_0/\Lambda^2)\,\sim \,10^{-14}$ in the range of interest 
$\muphi \sim \rm{TeV}$, $\Lambda \sim M_{GUT}$, $\Lambda_0 \sim M_{Pl}$, and can be safely ignored.  

\subsection{Sparticle spectrum}\label{subsec:spectrum}

Having identified superfields of the Standard Model with various components of the mesons $\Phi$ and $\Phi_U,$ we may now make 
parametric estimates for the soft masses obtained by gauginos, sfermions, and the gravitino in the supersymmetry-breaking vacuum.   

There are three possible contributions to the sfermion masses.  One contribution can come from a 
direct coupling to supersymmetry breaking.  This is the case for the composite first generation 
sfermions in $X_U$ that obtain a (large) mass from the Coleman-Weinberg potential,
\begin{equation}\label{eq:msf1}
V_{CW} \sim m_{CW}^2 |X_U|^2\;,\; m_{CW} \sim \sqrt{b} h^2 \mu.
\end{equation}
The composite second generation sfermions arising from $X$ have couplings to the supersymmetry breaking 
sector that are suppressed by the ratio $\t m_U / \Lambda$.  For $\t m_U/\Lambda \ll 1$, the second generation 
sfermions obtain only a negligibly small 
mass from the Coleman-Weinberg potential, even though they are composites!  
The gauginos and third generation do not have tree level couplings to the supersymmetry breaking fields.

The second contribution to the sfermion masses comes from gauge mediation.\footnote{A third possible contribution, which is incalculable, would come from
corrections to the canonical K\"ahler potential in the magnetic theory.  These can be expected to give contributions to soft masses of order $\mu^2/\Lambda$.  With
our choices of scales, such incalculable contributions are much smaller than the contributions from gauge mediation, and can be safely ignored.}
After weakly gauging, for example, an $SU(5)$ or $SU(3)_C\times SU(2)_L \times U(1)_Y$ subgroup of the 
global $SU(N_f-\Nct)$ symmetry as in (\ref{eq:embed}), the fields $\rho, \rhot, Z_U$, and $\ZUt$ 
will be charged under the Standard Model gauge group and act as messengers of  
supersymmetry breaking to the sparticle sector.  (We have seen in \S \ref{subsec:adj-dsb} and 
\S \ref{subsec:fermions} that the fields $Z$ and $\Zt$ can be decoupled from the supersymmetry breaking 
sector and be given very heavy masses of $\OO(\Lambda)$, so their interactions with the sparticle sector can be 
completely ignored.)  
The messenger masses may be computed from (\ref{eq:reducedFermion}) and (\ref{eq:bosons}); we refer the reader to \cite{Essig:2008kz} for the details.  
Very roughly, at leading order the fermionic components have masses $\sim h\mu$, while the bosonic 
components have masses $\sim 0, h\mu$, and $2h\mu$; the massless bosons will acquire a mass $\sim g_{SM} \mu$ when the flavor group is gauged.  
In the Standard Model embedding of (\ref{eq:embed}), we have $4 \times (\bf 5 +\bf \bar{5})$ messengers, so 
that gauge coupling unification is in principle possible (for a detailed discussion of unification in these models, see Appendix \ref{sec:unification}).

The gauge mediated two-loop contribution to the sfermion squared masses is parametrically given by
\be
m^2_{GM} \sim C \,\left(\f{g^2}{16\pi^2}\right)^2 \, \f{(hF_{X_U})^2}{M^2},
\ee
where $g$ is a Standard Model gauge coupling, $F_{X_U} \sim h \mu^2$ is the supersymmetry 
breaking $F$-term of the field $X_U$, and $M\sim h\mu$ is a typical messenger mass.   
We have neglected a sum over Dynkin indices and $\OO(1)$ numbers --- the precise 
expression is much more complicated and will not be needed for our purposes.  
The factor of $C$ counts the number of $Z_U$ and $\ZUt$ that are ${\bf 5}$'s or 
${\bf \bar{5}}$'s of the Standard Model $SU(5)$ gauge group. For Eq.~(\ref{eq:embed}) this is $C=2 \Nct$ (in the above example we have set $\Nct=1$), while for the model discussed in \S \ref{subsec:decoupling}, 
$C=3 \Nct$.
Schematically, the gauge-mediated contribution to sfermion soft masses is thus
\be\label{eq:soft-mass}
m_{GM} \sim \sqrt{C}\, \f{g^2}{16\pi^2} \, h\mu.
\ee

An interesting consequence of unifying flavor and supersymmetry breaking is that the Yukawa superpotential Eq.~(\ref{eq:Yuk}) gives matter-messenger couplings, because the latter also arise from the composite mesons. 
This can give a third possible contribution to the sfermion masses.  
Such matter-messenger mixings will be largest for the third generation:
\begin{equation}\label{eq:mess-matter}
W_{Yuk} \supset \f{1}{M_{\rm flavor}} (Q\t Q) H \Psi_3 + \f{1}{M_{\rm flavor}^2} (Q\t U Q) H \Psi_3 \,\rightarrow\, (\epsilon Z + \epsilon^2 Z_U)H \Psi_3 + \ldots
\end{equation}
Integrating out the messengers produces a negative one-loop contribution to the sfermion mass,
\begin{equation}\label{eq:extra}
\delta m^2 \sim -\frac{\epsilon^{2a}}{16 \pi^2} (h\mu)^2
\end{equation}
where $a=1$ or $2$ depending on whether the messengers come from $Z$ or $Z_U$. In the model of this section, 
supersymmetry breaking is produced by $X_U$ which couples predominantly to $Z_U$. Then (\ref{eq:extra}) is 
negligible compared to the gauge-mediated contribution. 
However, in \S \ref{sec:sols}, we will present realistic models where supersymmetry is broken by a linear combination of $X$ and $X_U$; in this case the coupling $ZH \Psi_3$ produces a negative contribution to the stop mass (\ref{eq:extra}) with $a=1$, which is of the same order of magnitude as the two-loop gauge mediated mass. Therefore messenger-matter mixings can significantly decrease the stop mass. Modifications of gauge mediation to include such mixings were studied in~\cite{Dine:1996xk}.

We next consider the gauginos, which receive a gauge mediated mass given in \cite{Essig:2008kz}.  
The mass must be proportional to the $R$-symmetry breaking, which 
is dominated by the spontaneous breaking from the vev of $\langle h X_U\rangle \sim \muphi/b$.   
Roughly,
\be
m_{\lambda_a} \, \sim \, C \,\f{g_a^2}{16\pi^2} \, \langle h X\rangle\,\sim \, 2\Nct \,g_a^2 \muphi\,,
\ee
where $g_a$, $a=1,2,3$, are the Standard Model $SU(3)_C$, $SU(2)_L$, and $U(1)_Y$ gauge couplings.
Notice that the $1/b$ factor in the the spontaneous $R$-symmetry breaking vev, $X \sim \muphi /b$, cancels the 
loop factor.  

The gauge mediated contribution to the sfermion and gaugino masses are in principle comparable if
\begin{equation}
\muphi \sim \mu/(16 \pi^2).
\end{equation}
Gauge mediated masses of $\OO(1~{\rm TeV})$ are obtained if (assuming $h\sim 1$ for now)
\be
\muphi \sim 1~{\rm TeV}, \;\;\; \sqrt{F} \sim \mu \sim \mathcal O (100 - 200\,\,{\rm TeV}), 
\ee
so that the direct supersymmetry breaking contribution from the Coleman-Weinberg potential to the first (and possibly second) generation sfermions is 
\be
m_{CW} \sim 10~{\rm TeV}.
\ee

A more detailed analysis reveals that the gauge mediated contribution to the colored sfermions in this simple model 
is larger than the gauge-mediated contribution to the gaugino masses.  
In the model of this section, where $C=2 \Nct$ (Eq.~(\ref{eq:embed})), setting the bino mass near its lower 
bound of $\sim 149$ GeV \cite{Aaltonen:2009tp,Meade:2009qv}, 
gives a stop mass of, very roughly, $\sim 4.5$ TeV for $\Nct=1$  and $\sim 3$ TeV for $\Nct=2$. 
For the model in \S \ref{subsec:decoupling}, 
$C=3 \Nct$, so that the stop mass is at least, very roughly, $\sim 3.5$ TeV for $\Nct=1$  and $\sim 2.5$ TeV for $\Nct=2$. 
This makes the model mildly tuned.  
However, as discussed above, in the more realistic model presented in \S \ref{subsec:decoupling}, the one-loop tachyonic contribution from messenger-matter mixing, Eq.~(\ref{eq:extra}), is effective in reducing the stop mass. This mechanism thus helps to avoid a hierarchy between gaugino and stop masses, that would otherwise be present.  Other ways to avoid this hierarchy would be to explore alternative 
classes of vacua (perhaps along the lines of \cite{Kom}), where
R-symmetry breaking comes about in a different way.

Finally, the gravitino mass in this theory is simply given by 
\be
m_{3/2} \sim \sqrt{\frac{N_f - \tilde N_c}{3}} \frac{h \mu^2}{M_P}.
\ee
For the low supersymmetry breaking scale considered here, the gravitino is light and has a mass of 
\be
m_{3/2} \sim 10~{\rm eV},
\ee
which makes it cosmologically quite safe \cite{Viel:2005qj}.

\subsection{Supersymmetric flavor}\label{subsec:comments1}

An essential challenge faced by single-sector models --- and, indeed, by all models of supersymmetry breaking and mediation --- is 
to generate a spectrum of soft masses compatible with observational constraints on flavor-changing neutral currents (FCNCs).  In 
general, the soft masses for squarks and sleptons explored in \S\ref{subsec:spectrum} are not diagonal in the same basis as the 
fermion mass matrix, leading to potentially prohibitive FCNCs.\footnote{In the single-sector models discussed in this paper, 
FCNCs do not only potentially originate from a misalignment of the fermion Yukawa matrices and the sfermion soft masses, but also 
from the fact that the Standard Model fermions couple directly to the messengers, because both are composite. 
Therefore, there are one-loop contributions to, for example, $K^0 - \bar{K}^0$ mixing from box 
diagrams containing messengers.  We will discuss these in \S \ref{app:messloops}, and find that they do not impose an important constraint 
on the models discussed in this paper, since they are suppressed by a loop factor and the large messenger mass.  
Furthermore, (\ref{eq:mess-matter}) (and similarly the other Yukawa superpotential terms) can generate 
FCNC's from box diagrams, but these are further suppressed by two or more powers of $\epsilon$ and therefore also negligible.} 
But the virtue of {\it calculable} models of single-sector supersymmetry breaking 
and flavor is that phenomenologically viable spectra may be related directly to microphysical parameters of the theory, and viable models 
may be found as a function of such parameters. In light of the potential soft terms discussed above, it is thus natural to consider what 
ranges of ultraviolet parameters in the adjoint model give rise to supersymmetric soft spectra compatible with experimental constraints. 

Absent any additional mechanism to generate alignment between the Yukawa matrices and sfermion soft masses, spectra compatible with 
FCNCs may arise from either approximate universality or decoupling. Universality --- for which the sfermion mass matrices are proportional 
to the identity --- suffices because the identity is diagonal in any basis, so that no sfermion mass mixing is generated when we rotate to the fermion mass eigenbasis. 
Although small deviations from universality are acceptable (and, indeed, inevitable given RG evolution of soft parameters to the weak 
scale), they must remain rather small compared to the overall scale of soft masses.

Decoupling, on the other hand, exploits the observation that sfermion contributions to FCNCs scale as high inverse powers of the sfermion 
mass, and vanish as the sfermion masses are taken to infinity. The size of the top Yukawa coupling implies that only the third generation of 
sfermions needs be near the weak scale to preserve the naturalness of weak-scale supersymmetry as a solution to the hierarchy problem. 
Fortunately, FCNC constraints are strongest for the first two generations of sfermions, so that flavor constraints and naturalness may be 
simultaneously satisfied by making the first two generations heavy while keeping the third generation light. This approach leads to ``more 
minimal'' \cite{Dimopoulos:1995mi,Cohen:1996vb} models with an inverse hierarchy of sfermion masses. In such scenarios, the masses of 
the first two generations of sfermions are constrained by the two-loop sfermion contribution to the stop mass, which renders the stop 
tachyonic when $m_{\tilde f_1}, m_{\tilde f_2} \gtrsim 20$ TeV unless the high-scale stop mass is unnaturally large \cite{AHM}.

In the models considered here, sfermions of the first two generations may acquire supersymmetry - breaking soft masses directly, while all three 
generations acquire universal gauge mediated soft masses. Barring additional superpotential terms mixing the 
mesons of the magnetic theory, these soft masses are all diagonal in the same basis as the non-diagonal Yukawa textures (\ref{simple}). If 
the gauge-mediated contributions are not more than a few TeV, the third generation of sfermions is light enough to 
roughly preserve the 
naturalness of electroweak symmetry breaking. It is then a question of how large the additional contributions to the first and second 
generations coming from $m_{CW}$ must be in order to avoid FCNCs. In general, both must be $\gtrsim 5$ TeV with some degree of 
degeneracy; a detailed treatment of FCNC and other constraints on the sfermion spectrum is contained in Appendix \ref{sec:FCNC}.

The great virtue of calculable single-sector models is that these flavor constraints may be related explicitly to the UV parameters of the 
theory.  In the limit $\t m_U/\Lambda \ll 1,$ only the first generation feels supersymmetry breaking directly. In general, such a spectrum --- with 
sfermions of the first generation much heavier than those of the second and third --- yields prohibitive contributions to FCNCs. However, if 
the coupling $h$ is sufficiently small, it is possible for such contributions to satisfy approximate universality given a certain degree of tuning. 
For larger values of $\t m_U/\Lambda$, both first and second generations obtain significant soft masses directly from supersymmetry breaking, 
realizing a calculable version of the ``more minimal'' scenario. This is perhaps the most natural spectrum of supersymmetry breaking in such 
theories, and (calculably) reminiscent of the dimensional hierarchy spectra in \cite{Luty:1998vr}. Finally, it is possible for all three 
generations to receive soft masses solely from direct gauge mediation if the chiral ring is extended slightly. These models naturally satisfy 
FCNC constraints via universality.

We will now detail these approaches in \S \ref{sec:sols}.

\section{Models}\label{sec:sols}

In light of the potential soft terms described in \S\ref{subsec:spectrum} and the supersymmetric flavor constraints outlined in \S
\ref{subsec:comments1} and Appendix A, let us now consider various limits of the adjoint theory that give rise to 
phenomenologically viable spectra. In \S
\ref{subsec:approx.}, we will consider the theory in \S \ref{subsec:adj-dsb}, which will give approximately universal 
sfermion masses; it involves the simple embedding discussed in \S \ref{subsec:fermions}, but requires 
some degree of tuning to satisfy FCNC constraints. In \S\ref{subsec:decoupling}, we will consider models with the familiar inverse hierarchy 
of soft masses; these models readily satisfy flavor constraints but entail a slightly less minimal embedding of Standard Model fields. In \S
\ref{subsec:exact}, we expand the chiral ring of the adjoint model of \S \ref{subsec:adj-dsb} 
to include theories where all three generations obtain universal masses 
from direct gauge mediation. In this case, the composite field that breaks supersymmetry is distinct from those giving rise to Standard Model 
generations, but all the ingredients of supersymmetry breaking, mediation, and flavor are contained within the same gauge sector.

\subsection{A model with approximate universality}\label{subsec:approx.}

We begin by exploring the simplest single-sector model that requires only the minimal Standard Model embedding of (\ref{eq:reps2}). 
Though admittedly not the most elegant model, this approach will illustrate some of the issues that will reappear in more elaborate 
alternatives.

In the limit $\t m_U / \Lambda \ll 1$, only the first generation feels supersymmetry breaking directly; the meson $\Phi$ in which 
the fields of the second generation are embedded remains approximately supersymmetric. Gauging the flavor symmetry then produces 
universal gauge-mediated masses for all three generations. From Eqs.~(\ref{eq:msf1}) and (\ref{eq:soft-mass}), 
these respective soft masses are
\begin{equation}\label{eq:sfermions}
m_{CW} \sim \frac{h}{4\pi}\,h\mu\;,\;m_{GM} \sim \frac{\alpha_g}{4\pi}\,h \mu\,,
\end{equation}
where $\alpha_g= g_{SM}^2/4\pi$.  
The first generation thus obtains a mass-squared of $m^2_{\t f_1} \sim m^2_{CW}+m^2_{GM}$ 
while the second generation obtains a mass-squared of only $m^2_{\t f_2} \sim m^2_{GM}$.  
For low sfermion masses where $m_{GM}\sim 3$ TeV, we need $m_{\t f_1}$ to be the 
same as $m_{\t f_2}$ within $\sim 2-5\%$ in order to avoid large FCNCs (see Appendix A).  
This requires the Coleman-Weinberg contribution to the first generation mass to be smaller than the gauge-mediated mass, 
which may be achieved only if $h \lesssim \alpha_g/4$.  
There is no reason for $h$ to be so small, but it is interesting that tuning a single dimensionless coupling 
can help solve the problem from FCNCs.  
In this case, the direct supersymmetry-breaking mass from the one-loop effective potential is much smaller than 
the gauge-mediated mass, and the spectrum looks like a very minor deviation from that 
of standard gauge mediation. 

One tension in the reasoning of the previous paragraph comes from the observation that $h \ll 1$ is in conflict with astrophysical constraints 
that imply a lower bound $h \gtrsim \mathcal O(1)$.  
Indeed, recall that in scenarios with a low scale of supersymmetry breaking and warm gravitino dark matter  the gravitino mass has 
an upper bound of $\sim 16$ eV, which translates into a bound on the supersymmetry-breaking scale of \cite{Viel:2005qj}
\begin{equation}
V_{min}^{1/4}= |\sqrt h\,\mu| \lesssim 260 \,{\rm TeV} \,.
\end{equation}
Fixing the stop mass in (\ref{eq:sfermions}) then gives a lower bound on $h$,
\begin{equation}
\sqrt h \gtrsim \frac{4\pi}{\alpha_g}\, \frac{m_{\t t}}{260\,{\rm TeV}} \sim \mathcal O(1)\,.
\end{equation}
Of course, this bound may be obviated by large entropy production at late times.  

Absent a cosmological solution, this tension may also be removed by the following simple modification. 
Let us allow two different $\mu$ parameters, $\mu_1 > \mu_2$,
\begin{equation}
W \supset -h \tr (\mu^2 \Phi_U) = -h \mu_1^2 \,\tr\,Y_U- h \mu_2^2 \,\tr\,X_U\,. 
\end{equation}
(Notice that nothing forbids such different $\mu$'s once the global symmetry group is explicitly broken by weakly gauging the 
Standard Model subgroup.) By the rank condition, the VEV of $\chi$ is set by the largest $\mu_1$,
$$
\langle \chi \t \chi \rangle = \mu_1\,.
$$
On the other hand, the supersymmetry-breaking scale is
$$
|W_{X_U}|=|h \mu_2^2|\,.
$$

In this more general setup, the direct and gauge-mediated masses become
\begin{equation}
m_{CW}\approx \frac{h}{4\pi}\,\frac{h \mu_2^2}{\mu_1}\;\;,\;\;m_{GM}\approx \frac{\alpha_g}{4\pi}\,\frac{h\mu_2^2}{\mu_1}\,.
\end{equation}
The upper bound on the scale of supersymmetry breaking from the astrophysical bound on the gravitino mass now does not limit $h$, but rather
\begin{equation}
\mu_1 \lesssim \frac{\alpha_g}{4\pi}\,\frac{(260\,{\rm TeV})^2}{m_{\t t}}\,.
\end{equation}
Then it is possible for $h$ to be small enough to satisfy approximate universality. Although the tuning of $h$ to accommodate FCNC constraints is somewhat arbitrary, it gives rise to a satisfactory spectrum of sfermions in the simplest embedding of Standard Model fields into the adjoint model.

\subsection{A model with decoupling}\label{subsec:decoupling}

A more familiar approach to viable single-sector supersymmetry breaking with a dimensional hierarchy is to adopt 
a decoupling solution in which the first- and second-generation sfermions are heavy. Indeed, this is the 
natural spectrum arising in adjoint models for finite values of $\t m_U / \Lambda.$

From the couplings in the superpotential  (\ref{eq:adj-Wmag4}), the supersymmetry-breaking sector induces a soft mass for the second generation at one loop of order
\begin{equation}
m_{CW,2} \sim \left(\frac{N_c-N_f}{g_U} \frac{m_U}{\Lambda} \right)\,\frac{h}{4\pi} \,h\mu,
\end{equation}
where the factor inside the brackets comes from the fact that the interaction between $\Phi$ and the magnetic quarks is proportional to 
$m_U/\Lambda$, and the second factor is the usual Coleman-Weinberg mass (\ref{eq:mcw}). Order-one numerical factors coming from the precise 
matching (\ref{eq:adj-params}) have been absorbed into $g_U$, and we have set $\Nct = 1$. 
Recall that $m_U$ and $g_U$ are the mass and cubic coupling of the 
adjoint field $U$ in the electric theory.

In our case, $(N_c - N_f) \sim \mathcal O(10)$ and $g_U$ can be made smaller than one. By taking $m_U/\Lambda$ small but finite (unlike 
the case $m_U/\Lambda \to 0$ of \S\ref{subsec:adj-dsb} and \S\ref{subsec:approx.}), it is possible to obtain
\begin{equation}\label{eq:largemU}
\frac{N_c-N_f}{g_U} \frac{m_U}{\Lambda} \sim \mathcal O(1)\,.
\end{equation}
For $h \sim \mathcal O(1)$, the direct supersymmetry breaking mass contribution is larger than the gauge mediated effect,
\begin{equation}
m_{CW,2} \sim \,\frac{h}{4\pi} \,h\mu > \frac{\alpha_g}{4\pi}\,h\mu
\end{equation}
and both first- and second-generation sfermions can be made much heavier than the stop.

There is, however, a small obstacle to this simple picture that needs to be overcome. 
From the superpotential (\ref{eq:adj-Wmag4}), the magnetic quarks $q, \t q$ only couple to the linear combination
\begin{equation}\label{eq:LC}
\frac{N_c-N_f}{g_U} \frac{m_U}{\Lambda} \Phi + \Phi_U
\end{equation}
which gets a mass from the one-loop Coleman-Weinberg potential
\begin{equation}\label{eq:Vcwsf}
V_{CW} \approx m_{CW}^2\,\tr\left[\left(\frac{N_c-N_f}{g_U} \frac{m_U}{\Lambda} X + X_U \right)^\dag\,\left(\frac{N_c-N_f}{g_U} \frac{m_U}{\Lambda} X + X_U \right)\right]\,.
\end{equation}
The orthogonal combination remains light. Therefore, at first glance it seems that the effect of 
increasing the coefficient $m_U/\Lambda$ is simply to redefine which scalar acquires a one-loop mass and 
which scalar receives a mass only from gauge mediation. At the level of the sfermion mass matrices, however, this would 
generate large off-diagonal elements strongly constrained by FCNCs; such mixings would require prohibitively large 
sfermion masses $\gtrsim 100$ TeV to evade flavor constraints.

We can solve this problem by noticing that if the first generation sfermions $({\bf 10}+ {\bf \bar 5})$ come from matrix elements $X_{U,ij}$ 
which are different from the matrix elements $X_{kl}$ containing the second generation, then (\ref{eq:Vcwsf}) will give independent masses 
to each of the Standard Model sfermions. In other words, both generations can come from the linear combination (\ref{eq:LC}) albeit from different matrix 
elements, and both then acquire comparable one-loop masses.

For this, we need to be able to have two different $({\bf 10}+ {\bf \bar 5})$ inside each meson. The minimal choice corresponds to
$$
N_f=17\;,\;N_c=33
$$
with the $SU(5)_{SM}$ embedding
$$
Q \sim {\bf 1} + [{\bf 1}+{\bf 5}+{\bf 5}+{\bf \bar 5}]\;,\;\t Q \sim {\bf 1} + [{\bf 1}+{\bf \bar 5}+{\bf \bar 5}+{\bf 5}]\,.
$$
Each of the mesons $X$ and $X_U$ contains two independent $({\bf 10}+ {\bf \bar 5})$'s, plus additional matter 
that is lifted by coupling it to spectator fields. The corresponding Standard Model generations are identified with orthogonal 
elements ${\bf 10}+ {\bf \bar 5}$.
To ensure that this happens, the superpotential coupling Eq.~(\ref{eq:additionalMatter}) of the spectators to the appropriate matrix elements can be enforced by an approximate discrete symmetry. For instance, we can consider a vector-like $\mathbb Z_2$, with charge assignments $Q \sim {\bf 1}_+ + [{\bf 1}_++{\bf 5}_++{\bf 5}_-+{\bf \bar 5}_+]$, opposite charges for $\t Q$, and with $U$ being odd. Introducing, in particular, ${\bf {\overline{ 10}}}_-$ and ${\bf 5}_-$ spectators, the ${\bf 10}_-+{\bf \bar 5}_-$ mesons are lifted. Only the ${\bf 10}_++{\bf \bar 5}_+$ from each $Q \t Q$ and $Q U \t Q$ survive --- and these come from different matrix elements since $U$ is odd. Notice that this discrete group commutes with the global symmetry group left unbroken by the $SU(5)_{SM}$ embedding. Also, since $U \to - U$ is not a symmetry in the presence of a $\Tr \,U^3$ superpotential, its coefficient $g_U$ has to be small in order for this analysis to be approximately correct. In practice, $g_U \lesssim \epsilon \sim \mathcal O(0.1)$ is required.
 
A fully realistic single-sector model satisfying the bounds from FCNCs is then possible, albeit with a slightly less minimal embedding of the Standard Model into composites of the strong dynamics. Let us consider a simple example.  Take the messenger scale to be
\be
M = h \mu \approx 250\;\text{TeV}\,.
\ee
Setting $h \sim \mathcal O(1)$, and $m_U/\Lambda \sim \mathcal O(0.01)$, the sfermion spectrum at the messenger scale is
\be
m_{\t f1} \approx 20\;\text{TeV}\;,\;m_{\t f2}\approx 15\;\text{TeV}\;,\;m_{\t f3} \sim 2.5\;\text{TeV}
\ee
The gaugino masses are
\be
m_\lambda \sim \mathcal O (100~{\rm GeV}-1\,\text{TeV})\;\text{for}\;\muphi \sim \mathcal O(100~{\rm GeV}-1\,\text{TeV}),
\ee
and the metastable vacuum is parametrically long-lived. In this class of models, the number of messengers is $6 \times ({\bf 5}+{\bf \bar 5})$ 
so that perturbative unification is not possible. It would be interesting to find a model that unifies and where the first two generation 
sfermions have decoupled to the multi-TeV scale.

As a final remark connecting with the discussion in \S \ref{subsec:more-general}, when (\ref{eq:largemU}) is satisfied the field breaking 
supersymmetry and $R$-symmetry is a linear combination of $\Phi$ and $\Phi_U$ with order one coefficients --- see Eq.~(\ref{eq:LC}). 
Turning on generic superpotential deformations $\Delta W_{el}=(Q \t Q)^n (Q U \t Q)^m$, the properties of the metastable vacuum will be 
fixed by only the largest linear and quadratic meson terms. These have to satisfy the stability conditions found in~\cite{Essig:2008kz}, while 
other terms play a subleading role. Therefore the metastable vacuum will exist and be long-lived for quite generic superpotential 
deformations.

\subsection{Composite models with direct gauge mediation}\label{subsec:exact}

So far we have found models where both composite generation sfermions acquire soft masses from direct couplings to the 
supersymmetry-breaking 
sector (see \S \ref{sec:simple} and \S\ref{subsec:decoupling}) or where the first generation gets a direct supersymmetry breaking mass, 
while the second predominantly obtains a mass from gauge-mediation  
(see \S\ref{subsec:approx.}). 
We saw that in order to satisfy FCNC constraints in the latter scenario,  the 
one-loop supersymmetry breaking mass must be considerably suppressed relative to 
the gauge-mediated masses. 

This limit suggests a slightly more general 
``single-sector'' scenario in which supersymmetry 
breaking still arises from strong dynamics of the $SU(N_c)$ gauge group, but all the soft masses 
come predominantly from gauge mediation. In this case, the flavor problem would be solved automatically due to 
the flavor-blindness of the gauge interactions, which produce universal sfermion masses. 
Though one might argue that this is no longer strictly a single-sector
theory --- the fields responsible for supersymmetry breaking only have
a highly suppressed coupling to the Standard
Model composite fermions --- such models still retain a pleasing
amount of compactness.
No new ingredients beyond 
the fields and interactions of the $SU(N_c)$ gauge theory are required, and all the messengers, supersymmetry-breaking fields, and Standard Model 
composites arise from the same dynamics. In this section, we present a simple deformation of the adjoint model discussed in 
\S \ref{sec:adjoint} possessing these 
properties.

Consider the adjoint model of \S \ref{sec:adjoint}, but allowing a $U^4$ term in the electric superpotential (the general $U^k$ case has been studied in~\cite{Kutasov:1995ss}),
\begin{equation}
W_{el}= \frac{1}{4}\,\frac{1}{\Lambda_U}\,\Tr U^4+\frac{g_U}{3} \,\Tr U^3 + \frac{m_U}{2} \,\Tr U^2
\end{equation}
The magnetic dual has gauge group $SU(\t N_c=3 N_f-N_c)$, $N_f$ magnetic quarks $(q, \t q)$, a magnetic adjoint $\t U$, and three gauge singlets
$$
M_1 = \t Q Q\;,\;M_2= \t Q U Q\;,\;M_3= \t Q U^2 Q\,.
$$
It will be useful to work in terms of canonically normalized mesons,
$$
\Phi_j = \frac{M_j}{\Lambda^j}\,,
$$
up to order one numerical constants from the K\"ahler potential as in Eq.~(\ref{eq:adj-mesons}).

Again, we will focus on the case $\t N_c=1$, for which the magnetic dual is a theory of weakly coupled hadrons with superpotential
\begin{equation}\label{eq:Wmag-direct}
W_{mag}= h\,\tr(\Phi_3 q \t q)+ \frac{2-N_c}{3}\, h^2 g_U\,\tr(\Phi_2 q \t q)+ h^2 \left(\frac{m_U}{\Lambda}
  + \frac{N_c^2-N_c-2}{9}\,h g_U^2 \right)\,\tr(\Phi_1 q \t q)
\end{equation}
where $h=\Lambda_U/\Lambda$. In the limit
\begin{equation}\label{eq:limit-gm}
hg_U \ll 1\;,\;h\frac{m_U}{\Lambda}\ll 1\,,
\end{equation}
the dimension 2 meson $\Phi_1$ and the dimension 3 meson $\Phi_2$ almost decouple from the rest of the low energy fields $(\Phi_3, q, \t q)$.

These fields $(\Phi_3, q, \t q)$ are then used to break supersymmetry in a by now familiar way. Adding the superpotential deformation
\begin{equation}
\Delta W_{el} \sim \frac{1}{\Lambda_0}\,(Q U^2 \t Q)+ \frac{1}{\Lambda_0^5}\,(Q U^2 \t Q)^2 \,,
\end{equation}
which in the magnetic theory becomes
\begin{equation}
\Delta W_{mag} \sim - h \mu^2 \,\tr\,\Phi_3+ h^2 \muphi \,\tr\,(\Phi_3)^2\,,
\end{equation}
breaks supersymmetry by the rank condition, creates a metastable vacuum at a distance $\sim 16 \pi^2 \muphi$ from the origin of $\Phi_3$ space, and breaks the $R$-symmetry both explicitly and spontaneously (the 
latter dominating).

The first and second SM generations are identified with $\Phi_2$ and $\Phi_1$ respectively, with the third generation being elementary. In the limit (\ref{eq:limit-gm}), none of the composite generations participate directly in the supersymmetry breaking. Therefore the sfermion soft squared masses come predominantly from gauge mediation, involving the supersymmetry breaking fields $(q, \t q)$ only at two loops. These contributions are flavor blind and hence there are no flavor problems since all the masses are universal.

It is quite surprising that calculable single-sector models exist where the composite soft masses come predominantly from direct gauge mediation. The gauge dynamics we have found is rich enough to provide marginal couplings ($g_U$ and $m_U/\Lambda$ in the example above) that control the strength of the direct supersymmetry breaking masses. It is possible to send these parameters to zero without changing the supersymmetry breaking scale and messenger masses. It would be interesting if this mechanism has an analog in single-sector models with gravity duals~\cite{Gherghetta,Landaupole,Shiu,toappear}.

\section{Concluding remarks}\label{sec:conclusions}

We have introduced and studied calculable models of single-sector supersymmetry breaking that have fully realistic Yukawa textures (implementing the dimensional hierarchy idea) and satisfy FCNC bounds, considerably improving  earlier constructions~\cite{Franco:2009wf}. The beauty of these constructions stems from the way in which the apparently intricate structure of the MSSM originates from a rather minimal, calculable gauge theory.

Our discussion focused primarily on a class of models based on SQCD with fundamental flavors and an adjoint. These theories possess composites of various dimensions, controlled by the adjoint superpotential, and exhibit a surprisingly wide range of interesting behaviors. In certain parametric limits they give rise to models in which first- and second-generation sfermions are heavy due to compositeness and decouple. Perhaps more unexpectedly, there are also models in this class where compositeness gives rise to realistic Yukawa matrices, but all sfermion masses come predominantly from gauge mediation and are thus universal. 

The parametric limits presented here represent a fraction of the possible single-sector models that may emerge from theories of SQCD with fundamental flavors and a rank-two tensor field. It would be useful to further explore the range of possible soft spectra that may be realized in such theories. Moreover, the models we have considered suffer somewhat from a large number of extra matter charged under the Standard Model gauge groups; it would certainly be interesting to find other examples of calculable theories with less unnecessary matter. 

Of course, such single-sector theories are but one approach (among many) for explaining the Standard Model flavor hierarchy. 
We conclude by comparing and contrasting the mechanism discussed in this paper with other 
explanations for the Yukawa hierarchies which exist in different classes of models.

\subsection{Comparison to other explanations}

The earliest class of explanations, and probably the best explored, use the Froggatt-Nielsen idea \cite{FN}.  
Here, one introduces a new $U(1)$ symmetry, $R$, broken by the vev of
a new scalar $\langle \phi_1 \rangle$ which has charge $+1$.  
One assumes that all of the Standard Model fermions are exactly massless in the limit that 
$R$ is unbroken -- that is, one assigns different charges to their left and right-handed
components.  Finally, one assumes the existence of some very heavy set of fermions (with various values of $R$) at a 
scale $\langle \phi_0 \rangle \gg \langle \phi_1 \rangle$, whose mass is set by the expectation value of
another $R$-neutral Higgs field $\phi_0$.   
By assigning appropriate charges under $R$ to the Standard Model fermions, one can then generate Yukawa couplings
suppressed by different powers of $\epsilon = \langle \phi_1 \rangle / \langle \phi_0 \rangle$.    Models which are broadly successful in accounting for flavor physics can
emerge from this framework.  Some of the most successful models have more than one small parameter.   The scales involved are not very tightly constrained by data, so
such models can account for observed physics and remain untestable in the foreseeable future.

An idea closely related to our own is to consider supersymmetric models where the MSSM generations interact with a strongly coupled superconformal field theory (at least over some range of
energies).  If the MSSM Yukawa couplings receive different anomalous dimensions, this can provide an explanation of Yukawa hierarchies \cite{NS}.  A recent exploration
of this idea appears in \cite{Poland}.  We note that this is very similar to our mechanism; here, the large anomalous dimension comes from the fact that the MSSM fields are
secretly composite and hence the Yukawa couplings are higher dimension operators above the compositeness scale $\Lambda$.  In addition, our mechanism correlates
this structure with the dynamics of supersymmetry breaking.  

A recent class of interesting, field-theoretic ideas appears in \cite{PS}.  These ``domino theories" are incompatible with conventional low-energy supersymmetry, but
are otherwise an economical proposal for generating realistic Yukawa textures. 

A very wide class of inter-related ideas uses the physics of extra dimensions:

\noindent
$\bullet$ In superstring compactifications, e.g. those of the heterotic string,
it is easy to find supersymmetric scenarios where the tree-level Yukawa couplings are related to topological invariants of the compactification
manifold.  These invariants often give some vanishing couplings, usually because the homology cycles on which some of the matter fields are localized do not intersect
with the Higgs or with the other matter field in the relevant Yukawa coupling.  In such a circumstance, the leading coupling is generated by world-sheet or space-time instanton effects, due to 
supersymmetric non-renormalization theorems.   
(The instanton is a non-local object in the internal dimensions, and can connect the disconnected homology cycles).
In a topology where only the top quark Yukawa is present at tree level, this can provide an attractive explanation for the rough features
of the fermion mass matrix.  See e.g. chapter 16 of  \cite{Ed} for an elementary introduction.  Note that this idea requires multiple parameters to match the observed spectrum, since
each instanton action is ${\it a ~priori}$ unrelated to the others; this idea also remains untestable until one reaches the compactification scale, which is typically $\sim M_{\rm GUT}$.
Many modern variants of this idea also exist in brane-world scenarios involving D-branes in Type II string theories.  For recent discussions in heterotic and type II models, see
\cite{Ovrut} and \cite{Dinst}, for instance.    Very recent work in the context of F-theory, where instantons do not play an important role in the attempts to explain flavor
physics, is summarized in \cite{F-theory}.

\noindent
$\bullet$  
In theories where the Standard Model gauge fields propagate in ``thick" branes (e.g. live in flat extra dimensions which are not excessively large), one 
can obtain Yukawa hierarchies
by localizing the matter fermions within these branes \cite{split} (see also \cite{Kaplan:2000av,Kaplan:2001ga}).  
In these split fermion scenarios, there are parameters governing both the location of the fermions (and the Higgs scalars),
and the thickness or form of their wavefunctions.  In many ways, this is similar to the first scenario above.
With a small set of such parameters, one can find acceptable scenarios.  These models can be (indirectly) testable at the TeV scale, but need not be
\cite{Hewett}.

\noindent
$\bullet$
In theories with warped (AdS-like) extra dimensions, with Standard Model gauge fields in the bulk, one can try to explain flavor by localizing fermions at different points along
the radial direction of AdS \cite{RS}.  Such theories are dual to large N gauge theories \cite{Juan}.  
Fields localized in the IR are composites of the CFT dynamics, while those localized in the UV are elementary fields external to the CFT. 
It can be of interest to have either an elementary Higgs (e.g.~in a supersymmetric scenario where supersymmetry is broken at the end of the warped throat geometry), or a 
composite Higgs (e.g.~in non-supersymmetric Randall-Sundrum scenarios).  In the former case, the fermions localized at the IR end of the geometry (which are highly
composite) will have the smallest Yukawa couplings, while in the latter case the highly composite fermions will have the largest Yukawa couplings.   In such scenarios, like
in the split fermion scenarios, there are again typically several parameters; they are now associated with the anomalous dimension of the CFT operator which couples the
Standard Model fermion to the large N CFT.    The non-supersymmetric scenarios of this sort are likely to be testable at the LHC due to the existence of charged, light KK
modes coming from the ${\rm TeV}$-scale end of the throat geometry.  In the supersymmetric scenarios this scale is considerably higher, since it is associated with
supersymmetry-breaking, and there may be no Standard Model charges visible at this scale in any case (since there is no need for the Standard Model gauge fields to have support
in the entire warped geometry).  In this general framework, there are in fact recent steps towards making holographic duals of models quite similar to the ones we have considered
\cite{Gherghetta,Landaupole,Shiu,toappear}.

In several of these cases, there are clear implications for the physics of grand unification.  In the Froggatt-Nielsen models, one must extend the GUT group by an additional $U(1)$ and add
new matter multiplets at a high scale.  This is not compatible with standard $SU(5)$ GUTs.  In the cases with split or warped localized fermions, one has the normal difficulties associated
with ``explaining'' unification as opposed to postulating it by tuning additional matter content (which is of course unnecessary in the MSSM). 
In particular in string theory realizations of the third scenario, it is challenging to avoid Landau poles, due to the large number of massive matter fields involving in typical constructions of
the observable sector and the large N CFT (see e.g. \S5\ of \cite{Landaupole}).
The case with instanton-suppressed
Yukawa couplings is naively compatible with unification, though it introduces new parameters and renders the apparent relations in 
e.g.~(\ref{simple}) somewhat ${\it ad~hoc}$.

The explanation of flavor in our single-sector models is most similar in spirit to the last extra-dimensional scenario we discussed, in the supersymmetric
case with an elementary Higgs and small couplings
for the highly composite fermions.  
The composites in our models are analyzed via Seiberg duality instead of using AdS/CFT duality, but both classes of models rely on compositeness to suppress Yukawa couplings.
We are close to having models which avoid Landau poles, but the pile-up of extra matter fields at the scale $\Lambda$ where the composite generations are generated remains
an obstacle to making models with honest, weakly-coupled unification.
Since our models involve at most one or two parameters in the flavor sector, they are quite competitive in terms of predictivity with all of the classes of scenarios
enumerated above. The correlation between soft-terms and Yukawa couplings, evident in most of the single-sector models (with at least one and often both of the first two generations having large
sparticle masses in most of the known classes of models), is a further prediction which is absent in the non-supersymmetric theories, in supersymmetric realizations of the Froggatt-Nielsen mechanism, and in 
the methods based on instanton calculus in supersymmetric string compactifications.

\subsubsection*{Note Added:}  
After this paper was published, we learned that the precise models studied here admit other
metastable vacua which are lower in energy than the vacua we focused on.  Then, a more detailed study 
is required to determine whether the lifetime of our vacua are sufficient to accommodate realistic cosmology.  
However, a very minor change to the models -- adjusting some of the electric quark masses by 
an O(1) factor -- removes this issue and leaves the rest of our
discussion unchanged.  For details, see \S4 of \cite{Behbahani:2010wh}.
We thank D. Green, A. Katz and Z. Komargodski
for bringing this issue to our attention.

\subsection*{Acknowledgements}

We thank S.~Dimopoulos for useful discussions, and O.~Aharony, A.~Katz, Z.~Komargodski, N.~Seiberg, D.~Shih, and T.~Tait 
for useful comments or questions.
N.C. is supported in part by the NSF GRFP and NSF grant PHY-0244728. 
R.E. and G.T. are supported by the US DOE under contract number DE-AC02-76SF00515 at SLAC.
S.F., S.K. and G.T. are supported in part by NSF grant PHY-05-51164 at the KITP.

\appendix

\section{Constraints from Flavor Changing Neutral Currents}\label{sec:FCNC}

As is often the case with theories of supersymmetry breaking, the sfermion mass matrix is generally not 
diagonal in the same basis as the fermion mass matrix. The GIM mechanism does not operate for such 
general squark masses, leading to potential flavor-changing neutral currents in conflict with experimental 
bounds. In order to make meaningful contact with experimental limits, we will parametrize the contributions to 
flavor changing neutral currents (FCNCs) following \cite{Gabbiani:1996hi} (for an up-to-date analysis of FCNCs see 
\cite{Altmannshofer:2009ne}).
 
In the single-sector models under consideration, the Yukawa matrices $\lambda_u, \lambda_d, \lambda_e$ are generated at the scale 
$M_{\text{flavor}}$ with textures (\ref{simple}) dictated by the scaling dimensions of different composite states (in the case of the first two 
generations) or elementary states (in the case of the third generation) of the UV theory. 
When supersymmetry is broken, the squarks and sleptons of 
the first, or the first two, generations may acquire supersymmetry-breaking soft masses directly, while all three generations acquire universal supersymmetry-breaking soft 
masses from gauge mediation. Barring additional superpotential terms mixing the mesons of the magnetic theory, these soft masses are all 
diagonal in the same basis as the non-diagonal Yukawa textures (\ref{simple}). 

To reach the physical mass eigenbasis, the fermion mass matrices $M^u = v_u \lambda_u,$ $M^d = v_d \lambda_d,$ and $M^e = v_d \lambda_e$ may be diagonalized by bi-unitary transformations
\be
V_L^u M^u V_R^{u\dagger} = D^u \\
V_L^d M^d V_R^{d\dagger} = D^d \\
V_L^e M^e V_R^{e \dagger} = D^e\,,
\ee
where, for example,  $D^u = \text{diag} \{ m_u, m_c, m_t \}.$ Likewise, we may write the $6 \times 6$ squark mass matrices $\tilde M^{u 2}, \tilde M^{d 2}, \tilde M^{e 2}$ as 
\begin{eqnarray} \label{rotate}
\tilde M^{x 2} = \left(\begin{array}{cc} 
\tilde M_{LL}^{x2} & \tilde M_{LR}^{x 2} \\
\tilde M_{RL}^{x2} & \tilde M_{RR}^{x2}
\end{array} \right) \,,
\end{eqnarray}
where $x = u, d, e$ and, for example, $\tilde M_{LL}^{u2}$ is the soft mass matrix for the squarks $u_L$ coming 
from the doublets $Q$, while $u_R$ are those coming from the singlets $\bar{u}$.  Both 
$\tilde M_{LL}^{x2}$ and $\tilde M_{RR}^{x2}$ are Hermitian and come directly from 
soft masses, while $\tilde M_{LR}^{x2}$ and $\tilde M_{RL}^{x2}$ come from the trilinear $A$-terms. 
We will henceforth concentrate on the case where $A$ terms are vanishingly small at 
the supersymmetry-breaking scale (they will be regenerated by RG flow, but still suppressed by a loop factor), so that $\tilde M_{LL}^{x2}$ and $\tilde M_{RR}^{x2}$ are the quantities of interest. For simplicity, we will also assume that $\tilde M_{LL}^{x2}$ and $\tilde M_{RR}^{x2}$ are identical.

Although the sfermion mass matrices $\tilde M_{LL}^{x2}, \tilde M_{RR}^{x2}$ are generated without off-diagonal elements, the transformation to the fermion mass eigenbasis (\ref{rotate}) also rotates the sfermions and produces mass mixings between different generations of order
\be
(\delta \tilde M_{MN}^{x2})_{ij} = \left( V_M^x \tilde M_{MN}^{x2} V_N^{x \dag} \right)_{ij}
\ee
where the $M,N$ refer to $L$ and $R$.
In the case where the off-diagonal terms in $\tilde M_{LL}^{q2}$ and $\tilde M_{RR}^{q2}$ are smaller than the diagonal ones (as they are in the models of interest) and the $V_{L,R}^x$ are close to the identity, it is conventional to parameterize FCNC constraints via bounds on the dimensionless quantities 
\be \label{delta}
(\delta^x_{MN})_{ij} = \frac{\left( V_M^x \tilde M_{MN}^{x2} V_N^{x \dag} \right)_{ij}}{\sqrt{\left( V_M^x \tilde M_{MN}^{x2} V_N^{x \dag}\right)_{ii} \left( V_M^x \tilde M_{MN}^{x2} V_N^{x \dag}\right)_{jj}}}.
\ee
 The $\delta_{ij}$ thus measure the relative size of the off-diagonal components in the 
sfermion mass matrices in a basis where the fermion mass matrices are diagonal.  
They can be constrained from measurements of e.g. $K^0 - \bar{K}^0$ or $D^0 - \bar{D}^0$ mixing 
and the rare decays $\mu\to e\gamma$ and $b \to s \gamma.$

\subsection{Constraints on single-sector models}

Relatively careful constraints on the sparticle spectrum may be placed on single-sector theories such as those considered here, owing to 
the fact that the Yukawa textures and soft masses are both specified by the dynamics. This allows the degree of alignment between fermion 
and sfermion masses to be quantified, thereby ameliorating more conservative bounds on arbitrary mass matrices. Here we will place 
bounds on first- and second-generation sfermion masses for flavor models involving a Yukawa texture of the form (\ref{simple}). These 
constraints are germane to the single-sector models developed above, but also pertain to other flavor models with similar textures.

Constraints for FCNCs are by far the strongest on the down quark sector, owing to relatively tight limits on the $K_L - K_S$ mass difference. 
As such, we will focus here on bounds arising from the down sector, under the assumption that the sfermion masses in all three sectors will 
be approximately similar; bounds on the up quark and lepton sector provide considerably weaker constraints on the soft spectrum.

For simplicity, we consider a Yukawa texture of the form
\be
\lambda_d \simeq \left( \begin{array}{ccc}
\epsilon^4 & 2 \epsilon^3 & \frac{1}{4} \epsilon^2 \\
2 \epsilon^3 & 3 \epsilon^2 & \epsilon      \\
\frac{1}{4} \epsilon^2 & \epsilon    & \frac{1}{4} 
\end{array}
\right)\,,
\ee
where we have chosen the numerical coefficients to give us nonzero eigenvalues approximately reproducing the down-sector quark 
masses when $\epsilon \sim 0.1,$ $\tan \beta \sim 14,$ and $v = 246$ GeV. This gives us down, strange, and bottom masses 3 MeV, 152 
MeV, and 5 GeV, which are close to reality and give realistic FCNC bounds.  Naturalness dictates that the stop mass should not be much 
heavier than $1-2$ TeV, which sets the rough scale of gauge-mediated contributions to all three generations 
(in some, but not all, of the models we consider, the stop mass at the high scale cannot be much less than a few TeV --- see \S \ref{subsec:spectrum}).  
When this is the only source of supersymmetry breaking, (\ref{delta}) is 
always diagonal and FCNCs are negligible. However, in addition to the gauge-mediated contribution, the first and second generation 
squarks and sleptons may obtain additional soft masses directly from supersymmetry-breaking, leading to an inverse hierarchy. The size of 
additional contributions to the soft masses $m_{\tilde f_1}, m_{\tilde f_2}$ of the first two generations is then constrained by FCNCs.

The FCNC constraints are strongest for the parameter $(\delta^d)_{12},$ which parameterizes mixing of the first and second generation down-type squarks and is constrained by $K^0 - \bar{K}^0$ mixing; the bound is approximately $(\delta^d)_{12} \leq 2.5 \times 10^{-3} \frac{\sqrt{m_{\tilde f_1} m_{\tilde f_2}}}{500}$ for $m_{\tilde g}^2 \simeq 0.3 m_{\tilde f_1} m_{\tilde f_2}$ (and weakens with increasing gluino mass). The constraints on first- and second-generation mixing in the up quark sector from $D^0 - \bar{D}^0$ are weaker by roughly a factor of 2, while the constraints on the lepton sector from $\mu \to e \gamma$ are weaker still. We may also constrain the matrix elements $\delta^d_{13}$ from $B^0 - \bar{B}^0$ mixing and  $\delta^d_{23}$ from the rare process $b \to s \gamma,$ though again these constraints prove far weaker than those arising from $K^0 - \bar{K}^0$ mixing.  

We also note that in the single sector models of \S \ref{sec:sols}, the Standard Model fermions couple directly to the 
messengers, because both are composite. 
Therefore there are one-loop contributions to, for example, $K^0 - \bar{K}^0$ mixing from box 
diagrams containing messengers.  We will discuss these in \S \ref{app:messloops}.

\subsection{Constraints from $K^0 - \bar{K}^0$}

In order to constrain the possible values of $m_{\tilde f_1}$ and $m_{\tilde f_2}$ via the parameters $(\delta^d_{LL})_{12}$ and $(\delta^d_{RR})_{12}$, we can compute their contribution to the $K_L - K_S$ mass difference $\Delta m_K.$ This difference has been measured within excellent precision to be very nearly $\Delta m_K = (3.483 \pm 0.006) \times 10^{-12}$ MeV \cite{Amsler:2008zzb}. There are Standard Model contributions to this quantity that parametrically fall within the measured value, but depend on hadronic uncertainties to an extent that the full contribution is unknown. Thus we can take as our constraint the requirement that our contribution to $\Delta m_K$ does not exceed (in magnitude) the measured value.  We can extract the contribution to $\Delta m_K$ from squark mixing from \cite{Gabbiani:1996hi}. These contributions depend on the gluino mass $m_{\tilde g}$ and the squark masses $m_{\tilde f_1}, m_{\tilde f_2}$ via the mixings $(\delta^d_{MN})_{12}$ for $M, N = L, R.$ We will assume in our case that the $LR$ and $RL$ contributions are negligible and that $\delta_{LL} \simeq \delta_{RR}$, which is fairly accurate even when the Yukawa matrices are not entirely symmetric. This leads to by far the strongest constraints on the sfermion mass spectrum, as shown in Fig.~\ref{fig:1}.

\subsection{Constraints from $K^0 - \bar{K}^0$ from messenger loops}\label{app:messloops}

In the single sector models of \S \ref{sec:sols}, the Standard Model fermions couple directly to the messengers, 
because both are composite. Therefore there are one-loop contributions to, for example, 
$K^0 - \bar{K}^0$ mixing from box diagrams containing messengers.\footnote{We thank 
O. Aharony for bringing up this possibility.}

The dimension six operator induced by the messengers is of order
\begin{equation}\label{eq:box-mess}
H_{\Delta S=2} \sim \frac{1}{16 \pi^2 \mu^2}\,\bar d s\, \bar d s\,.
\end{equation}
Recall that
$$
\langle K^0|\bar d s\, \bar d s| \overline{K^0} \rangle \sim m_K f_K^2 = (497\;{\rm MeV})\;(160\;{\rm MeV})^2\,.
$$
Then imposing
\begin{equation}
\Delta m_K \,\sim \,{\rm Re} \langle K^0|H_{\Delta S=2} \overline{K^0} \rangle \approx 3.5\,\times\,10^{-12}\,{\rm MeV}
\end{equation}
on Eq.~(\ref{eq:box-mess}), gives a lower bound on the supersymmetry breaking scale,
\begin{equation}
\mu \gtrsim 160\;{\rm TeV}\,.
\end{equation}

This constraint can be accommodated in our models. It is interesting that FCNCs place a lower bound on the scale of supersymmetry breaking.

\subsection{Constraints from other processes: $B^0 - \bar{B}^0,$ $D^0 - \bar{D}^0,$ $b \to s \gamma,$ and $\mu \to e \gamma$}

The mixings $(\delta^d_{MN})_{13}$ may similarly be constrained by $B^0 - \bar{B}^0$ mixing from their contribution to 
$\Delta m_B = (3.337 \pm 0.033) \times 10^{-10}$ MeV \cite{Amsler:2008zzb}. The calculation is essentially identical to that
 of the previous case, with the replacements $m_K \to m_B,$ $m_s \to m_b$, $f_K \to f_B,$ and $m_{\tilde f_2} \to m_{\tilde f_3}.$ The 
 resulting constraint is much weaker than that from $K^0 - \bar{K}^0$.

We may constrain mixing between the second and third generations via the rare decay $b \to s \gamma$, using the gluino-mediated 
contribution in  \cite{Gabbiani:1996hi}. In this case, we require that our contribution not exceed the measured branching ratio $BR(b \to s 
\gamma) = (3.52 \pm 0.23 \pm 0.09) \times 10^{-4}$ \cite{Barberio:2008fa}. The branching ratio is a strong function of squark mass, and is 
satisfied readily for squark masses above 1 TeV. 

 Although we have focused here on the down sector, similar constraints on $(\delta^u)_{12}$ and $(\delta^e)_{12}$ arise from $D^0 - 
 \bar{D}^0$ mixing and the rare decay $\mu \to e \gamma$, respectively. Assuming the soft masses for all three sectors are parametrically 
 similar, these constraints are generally weaker than those considered above, so we do not show them explicitly. 

\begin{figure}[!t]
\begin{center}     
\includegraphics[width=.45\textwidth]{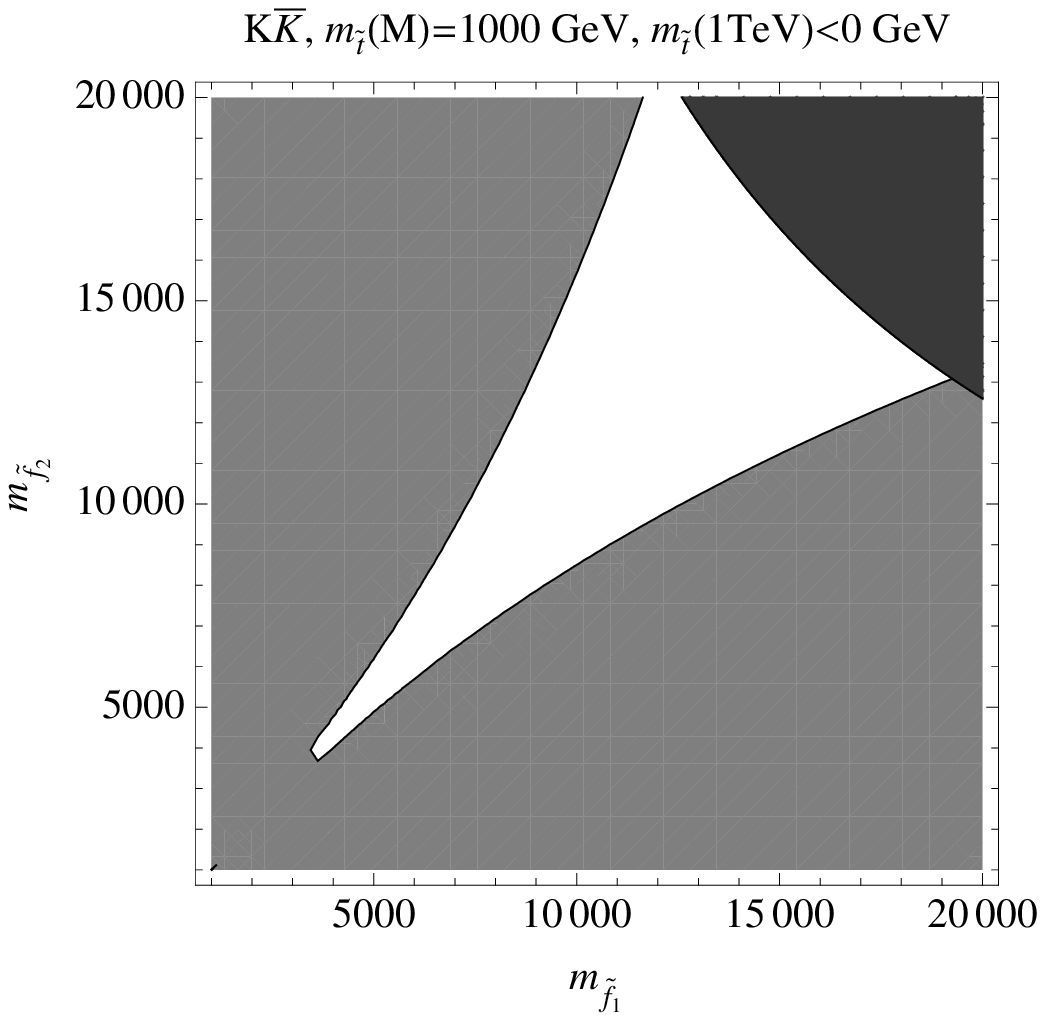}
\includegraphics[width=.45\textwidth]{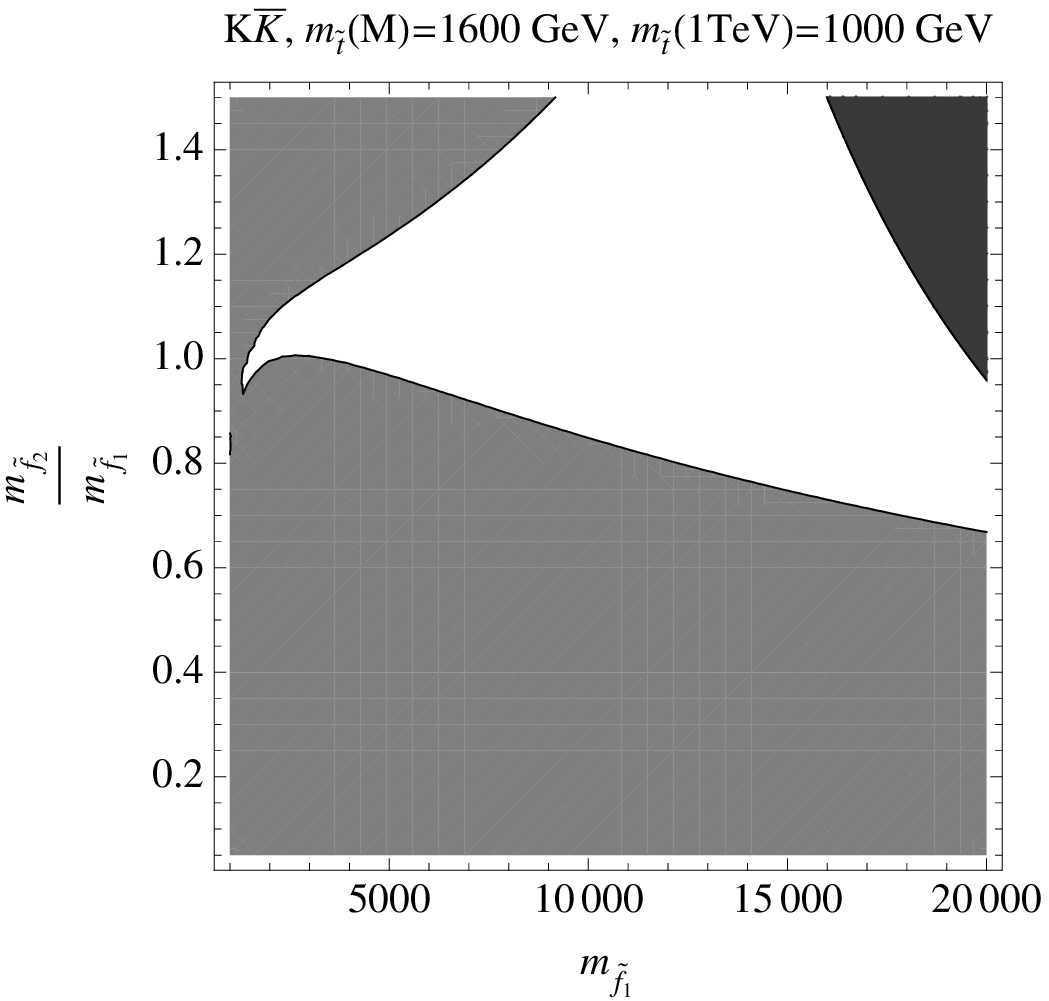}
\caption{Constraints on first and second generation sfermion masses. 
Light gray region ruled out by $K-\bar{K}$-mixing. 
(a) Dark gray region ruled out by tachyonic stops at the weak scale 
$(m_{\t t}(1 \,\rm{TeV})<0$).  
We assumed $m_{\t g}=500$ GeV, $m_{\t t}(100 \,\rm{TeV})=1$ TeV.  Note that the 
stop mass constraint disappears completely for the $m_{\t f_1}$ and $m_{\t f_2}$ mass range shown 
when $m_{\t t}(100 \,\rm{TeV})\gtrsim 1.6$ TeV.  
(b) Dark gray region ruled out by stops being too light at the weak scale 
to give a Higgs mass above LEP limits $(m_{\t t}(1 \,\rm{TeV})\lesssim 1000$ GeV, assuming the trilinear coupling is negligible).  
We assumed $m_{\t g}=500$ GeV, $m_{\t t}(100 \,\rm{TeV})=1.6$ TeV. }
\label{fig:1}
\end{center}
\end{figure}

\subsection{Constraints from tachyonic stop mass}

Finally, we can take into account the upper bound placed on squark masses by the desire for a positive stop mass at the weak scale. As 
noted in \cite{AHM}, overly large masses for the first and second-generation squarks can drive the stop mass negative via their two-loop 
contribution to the stop mass RG. We can place a conservative bound on the masses of first- and second-generation squarks by just 
considering the interplay between one-loop gaugino contributions and two-loop squark contributions to the stop soft mass. In particular, we 
will ignore the contribution from the top Yukawa, which can drive the stop mass more negative. We will also ignore the running of the first 
and second generation squark masses, which is (verifiably) negligible. In this simplified case, we can solve the RGE for the stop mass 
analytically to find \cite{AHM}
\begin{eqnarray}
m_{\tilde t}^2(\mu) \simeq m_{\tilde t}^2(\Lambda) + \sum_i \frac{2}{b_i} (M_i^2(\Lambda) - M_i^2(\mu)) C_i^{\tilde t} \nonumber \\
- 32 \tilde m_{1,2}^2 \sum_{i} \frac{1}{2 b_i} \left(\frac{g_i(\Lambda)^2}{16 \pi^2} - \frac{g_i(\mu')^2}{16 \pi^2} \right)  C_i^{\tilde t}
\end{eqnarray}
where $\tilde t$ can refer to $\tilde t_L$ or $\tilde t_R$ with appropriate choice of Casimirs (the stronger bound is on $\tilde t_L$), $i = 1,2,3,$ 
$b_i$ and $C_i$ are the usual GUT-normalized $\beta$ function parameters and Casimirs respectively, $\tilde m_{1,2}^2$ are the mean 
squark masses, $\mu$ is the low scale (taken to be 1 TeV), $\mu'$ is the scale where the heavy squarks decouple (taken to be 10 TeV), and 
$\Lambda$ is the scale where supersymmetry is broken and RG flow commences (taken to be 100 TeV). We also take $M_i = g_i^2 M_0,$ where 
$M_0 \sim \mu_\phi$ is the unified gaugino mass. 

We may use the running of the stop mass to place two potential constraints on the masses of first- and second-generation sfermions. A 
weak constraint is the requirement that the stop retain a positive mass-squared at the weak scale; a stronger constraint is that the stop mass 
remain large enough at the weak scale ($\sim 1$ TeV, neglecting the stop trilinear coupling \cite{Essig:2007vq}) to 
lift the Higgs mass above LEP limits. Aspects of both constraints are shown in Fig.~\ref{fig:1}.

\section{Unification}
\label{sec:unification}

As is often the case for theories involving additional multiplets charged under the Standard Model, it is natural to consider whether the perturbative unification of Standard Model gauge couplings may be preserved and low-scale Landau poles avoided.  Indeed, many models of metastable supersymmetry breaking suffer from the ubiquitous intermediate-scale Landau pole for the Standard Model gauge group. However, here it may be marginally possible to achieve unification at the GUT scale $\sim 10^{16}~ {\rm 
GeV}$.

Here we briefly recall the standard analysis of how extra $SU(5)$ multiplets affect the running of the gauge coupling.
The relevant formula, found in e.g. \S2\ of \cite{Giudice}, is that 
\be
\delta{\alpha_{\rm GUT}}^{-1} = -\frac{N}{2\pi} {\rm log}\left( \frac{M_{\rm GUT}}{M} \right)
\ee
where
\be
N = \sum_{i=1}^{K} n_i
\ee
is the sum of the Dynkin indices $n_i$ of the K extra $SU(5)$-charged matter multiplets.   So each ${\bf 5}$ or ${\bf \bar 5}$
contributes 1 to the sum, each ${\bf 10}$ contributes 3, each ${\bf 15}$ contributes 7, and each ${\bf 24}$ contributes 10.

The $4 \times ({\bf 5 + \overline{5}})$ messengers we have at the $\sim 100 ~{\rm TeV}$ scale, in our ``best" models, 
is a safe number to preserve
perturbativity of $\alpha_{\rm GUT}$, in absence of additional SU(5) charges at higher scales below $M_{\rm GUT}$. 
However, we have a large amount of additional matter at the scales
$\Lambda^2 / \Lambda_0$ and $\lambda \Lambda$.  Even under the assumption that $\Lambda \sim M_{\rm GUT}$ and
we can ignore running due to the latter, the states at $\Lambda^2/\Lambda_0$ will contribute a total Dynkin index 
given by summing over the representations in brackets in (\ref{eq:reps2}), multiplied by 2 (to include the ``spectators" they
pair with).  The total $N$ just from (\ref{eq:reps2}) is 40, and makes it somewhat challenging
to achieve unification before hitting a Landau pole, unless one pushes $\Lambda_0$ dangerously close to $M_{\rm GUT}$ or a larger Yukawa coupling is used.

It is important to remark that the non-spectator extra states are composites, which will in fact deconfine around the scale $\Lambda$.  Such composites will clearly contribute
differently to running at energies above $\Lambda$ (where we should use the electric description and count electric quark messengers), and it is conceivable that in some models this would vitiate the large threshold from encountering this plethora of states --- this has played a crucial role in the ideas of \cite{khoze}.  However, in our concrete models even the electric ``messenger index'' would be
quite large.  In addition, the precise
contribution in the energy regime around $\Lambda \sim M_{\rm GUT}$ does not seem easily calculable, and is naively quite significant.  Thus, although in our construction we have suceeded in pushing the Landau pole to very high scales, comparable to $M_{\rm GUT}$, it would also be interesting to find models where this problem is completely solved --- perhaps along the lines of~\cite{khoze}.

\section{Models with less extra matter}

\label{section_SM_reps}

Generically, the class of models discussed in this paper exhibit a proliferation of charged matter coming from 
$X$ and $X_U$. On one hand, this fact is an aesthetic nuisance since the corresponding masses, arising 
from cubic couplings in the electric theory, are naturally close to the high compositeness
scale $\Lambda$.  More importantly, as discussed in appendix B, these states affect the RG running 
at very high energies, making perturbative unification challenging. In addition, the models contain a large 
number of messengers in the $(\rho,Z_U)$ sector. These fields have masses $\sim 100$ TeV, and thus 
affect the running of couplings more dramatically. In certain cases, like the one in \S\ref{subsection_example} 
and the two composite generation example in \cite{Franco:2009wf}, these states lead to Landau poles below the GUT scale.

Throughout the paper, we have adopted an $SU(5)$ notation, mainly as a practical way of simplifying 
the group theory calculations, with the understanding that $SU(3)_C\times SU(2)_L\times U(1)_Y$ quantum 
numbers could be easily re-introduced at any step. In the absence of Landau poles, a physical 
consequence of the entire field content (except the two light Higgs doublets) fitting into $SU(5)$ 
representations is unification. In this section we explore what happens if we build models 
dropping the $SU(5)$ condition. We will see that both the amount of extra matter in $X$ and $X_U$ 
and the number of messengers is substantially reduced. 

We illustrate our ideas with the adjoint model 
of \S\ref{sec:adjoint}. The minimal model corresponds to taking $N_c=15$, $N_f=8$ and 
embedding the $SU(3)_C\times SU(2)_L\times U(1)_Y$ into $SU(8)$ according to
\begin{equation}
\begin{array}{cclcc}
Q & \sim & [({\bf 3},{\bf 1})_{x-1/3}+({\bf 1},{\bf 2})_{x-1/2}+({\bf 1},{\bf 1})_{x-1}+({\bf 1},{\bf 1})_{x}] & + & ({\bf 1},{\bf 1})_0 \\
\t Q & \sim & [({\bf \bar{3}},{\bf 1})_{1/3-x}+({\bf 1},{\bf 2})_{1/2-x}+({\bf 1},{\bf 1})_{1-x}+({\bf 1},{\bf 1})_{-x}] & + & ({\bf 1},{\bf 1})_0
\end{array}
\end{equation}
The parameter $x$ is fixed by imposing ${\rm Tr}\,(Y m^2)=0$, so that no FI term for $U(1)_Y$ is generated after integrating out the messengers. $X$ and $X_U$ decompose as

\begin{equation}
\begin{array}{c}
\left( ({\bf 3},{\bf 2})_{1/6} + ({\bf \bar{3}},{\bf 1})_{1/3} + ({\bf 3},{\bf 1})_{-2/3} + ({\bf 1},{\bf 2})_{-1/2} + ({\bf 1},{\bf 1})_{1}\right)
\\

+\left[({\bf 8},{\bf 1})_{0} + ({\bf \bar{3}},{\bf 2})_{-1/6}+ ({\bf 3},{\bf 1})_{2/3}+({\bf 3},{\bf 1})_{-1/3} +({\bf 1},{\bf 3})_{0} \right.
\\
+\left. 2 \times ({\bf 1},{\bf 2})_{1/2}+({\bf 1},{\bf 2})_{-1/2}+({\bf 1},{\bf 1})_{1}+4 \times ({\bf 1},{\bf 1})_{0} \right]
\end{array}
\label{Xi_SM_reps}
\end{equation}
namely, a full Standard Model generation plus additional matter, shown in square brackets. We see that the amount of extra 
matter in $X$ and $X_U$ has been reduced to less than a third of that in (\ref{eq:reps2}). $x$ naturally drops 
out from (\ref{Xi_SM_reps}), since it comes with opposite signs in the corresponding $Q$ and $\tilde{Q}$ entries.

Let us now focus on the messengers coming from the $(\rho,Z_U)$ sector. Their hypercharges do 
depend on the value of $x$. Interestingly, setting $x=0$ we can form a ${\bf 5}$ of $SU(5)$ by 
combining the $({\bf 3},{\bf 1})_{1/3}$ from $\tilde{\rho}$ and the $({\bf 1},{\bf 2})_{-1/2}$ from 
$\rho$ (and similarly for ${\bf \bar{5}}$ and $Z_U$ and $\tilde{Z}_U$). In this case, the messengers become   
\be
2 \times \left [({\bf 5}+{\bf \bar{5}}) + (({\bf 1},{\bf 1})_1+({\bf 1},{\bf 1})_{-1})+2 \times ({\bf 1},{\bf 1})_0 \right] 
\ee
where we have used a hybrid $SU(5)$-Standard Model notation to emphasize that the entire low energy 
spectrum is in full $SU(5)$ representations modulo two $Y=\pm 1$ pairs. The number of 
charged messengers is also reduced, by approximately a factor of $1/2$, with respect to the 
example in \S\ref{sec:adjoint}, ameliorating the Landau pole problem discussed in Appendix \ref{sec:unification}.


\end{document}